\documentclass[fleqn,usenatbib]{mnras}

\usepackage{newtxtext,newtxmath}

\usepackage[T1]{fontenc}
\usepackage{ae,aecompl}

\usepackage{graphicx}	
\usepackage{amsmath}	
\usepackage{amssymb}	
\usepackage{booktabs,dcolumn}
\usepackage{threeparttable}

\newcommand{\quotes}[1]{``#1''}
\newcommand{\arcsecs}[1]{\prime \prime}

\title[Vibrationally excited HC$_3$N  emission in NGC\,1068]{Vibrationally excited HC$_3$N emission in NGC\,1068: Tracing the recent star formation in the starburst ring.
}

\author[F. Rico-Villas  et al.]{
F. Rico-Villas$^{1}$\thanks{E-mail: fernando.rico@cab.inta-csic.es},
J. Mart\'in-Pintado$^{1}$, 
E. Gonz\'alez-Alfonso$^{2}$,
V. M. Rivilla$^{1,3}$,
\newauthor
S. Mart\'in$^{4, 5}$,
S. Garc\'ia-Burillo$^{6}$,
I. Jim\'enez-Serra$^{1}$,
M. S\'anchez-Garc\'ia$^{1}$
\\
$^{1}$Centro de Astrobiolog\'ia (CSIC-INTA). Ctra de Ajalvir, km. 4, Torrej\'on de Ardoz, 28850, Madrid, Spain\\
$^{2}$Universidad de Alcal\'a, Departamento de F\'isica y Matem\'aticas, Campus Universitario, Alcal\'a de Henares, 28871, Madrid, Spain\\
$^{3}$INAF-Osservatorio Astrofisico di Arcetri, Largo Enrico Fermi 5, 50125, Florence, Italy\\
$^{4}$European Southern Observatory, Alonso de C\'ordova, 3107, Vitacura, Santiago 763-0355, Chile\\
$^{5}$Joint ALMA Observatory, Alonso de C\'ordova, 3107, Vitacura, Santiago 763-0355, Chile\\
$^{6}$Observatorio Astron\'omico Nacional (OAN-IGN) - Observatorio  de Madrid, Alfonso XII, 3, 28014, Marid, Spain\\
}

\date{Accepted 2021 January 19. Received 2020 November 17; in original form 2020 August 10}

\pubyear{2020}

\begin{document}
\label{firstpage}
\pagerange{\pageref{firstpage}--\pageref{lastpage}}
\maketitle

\begin{abstract}


Using ALMA data, we have studied the HC$_3$N and continuum emission in the starburst pseudo-ring (SB ring) and the circumnuclear disc (CND) of the SB/AGN composite galaxy NGC\,1068. We have detected emission from vibrationally excited HC$_3$N (HC$_3$N*) only towards one star-forming region of the SB ring.
Remarkably, HC$_3$N* was not detected towards the CND despite its large HC$_3$N $v=0$ column density.  From  LTE and non-LTE modelling of HC$_3$N*, we obtained a dust temperature of $ T_\text{dust} \sim 250$\,K  and a density of $n_{\text{H}_2}=6\times10^5$\,cm$^{-3}$ for this star-forming region. The estimated IR luminosity of $5.8\times10^8$\,L$_\odot$ is typical of  proto-Super Star Clusters (proto-SSC) observed in the SB galaxy NGC\,253. We use the  continuum emissions at $147$\,GHz and $350$\,GHz, along with CO and Pa\,$\alpha$, to estimate the ages of other $14$\,SSCs in the SB ring.  
We find the SSCs  to be associated with the region connecting the nuclear bar with the SB ring, 
supporting  the inflow scenario.
For the CND, our analysis yields $T_\text{dust} \leqslant 100$\,K and $n_{\text{H}_2}\sim(3-6)\times10^5$\,cm$^{-3}$.
The very different dust temperatures found for the CND and the proto-SSC indicates that, while the  dust  in  the  proto-SSC  is  being  efficiently  heated from the inside by the radiation from massive proto-stars, the CND is being heated externally by the AGN, which in the IR optically thin case can only heat the dust to $56$\,K.
We discuss the implications of the non-detection of HC$_3$N* near the luminous AGN in NGC\,1068 on the interpretation of the  HC$_3$N* emission observed in the SB/AGN composite galaxies NGC\,4418 and Arp\,220.
\end{abstract}




\section{Introduction}
 In many galaxies, both active galactic nucleus (AGN) and star formation (SF) activity contribute to a significant fraction of the total galaxy luminosity \citep{Genzel1998}. Discerning how much each one contributes to the galaxy luminosity can tell us the physical processes heating the gas and dust and thus evaluate their associated radiative and kinematic feedback in the context of galaxy evolution. In addition, establishing which is the dominating heating mechanism (AGN or SF) in extremely obscured nuclei remains a key problem in extragalatic astrophysics \citep{Martin2016}. 
 
 Due to the high extinction in these environments, observations in the IR or at shorter wavelengths only detect the  outermost surface of the optically thick regions, where column densities of molecular gas can reach values of up to or even beyond $N(\text{H}_2)=10^{25}$\,cm$^{-2}$.
With such large column densities, the AGN  becomes Compton thick, preventing its identification even in the X-rays \cite[e.g.][]{GA2012, Lusso2013, Costagliola2013}.

In the last years it has been proposed that emission from vibrationally excited molecules such as HCN or HC$_3$N in highly obscured galactic nuclei  can be used as tracers of the nuclear activity in these obscured galaxy nuclei \citep{Sakamoto2010, Costagliola2010, Martin2011, Martin2016, Aalto2015a, Imanishi2016c, Imanishi2019, GA19}.
As the radiation from the obscured heating source is reprocessed by the dust into the IR, peaking in the mid-IR ($10-50$\,$\mu$m), the IR photons efficiently pump these molecules into their vibrational excited states. On the other hand, the rotational levels from the vibrationally excited molecules emit radiation in the cm to sub-mm wavelength range,  unaffected by dust extinction, making them powerful tools to probe the energetic processes taking place in the innermost regions of  heavily obscured galactic nuclei.

In particular, vibrationally excited HC$_3$N (hereafter, HC$_3$N*) is very well suited for tracing energetic processes in highly obscured regions. In the Milky Way (MW), its emission is associated with warm, dense UV shielded and very compact regions around massive star-forming regions, named Hot Cores \citep[HCs, e.g.][]{Goldsmith1982, Wyrowski1999, deVicente2000, deVicente2002, Jesus2005}, where HC$_3$N* is tracing massive star formation in the early protostellar phases \citep[hot core and circumstellar phases;][]{Izaskun2009}. However, HC$_3$N* emission is not detected in the circumnuclear disc (CND) surrounding the supermassive black hole (SMBH) of the MW since HC$_3$N is photodissociated  by UV radiation \citep{Costagliola2010, Martin2012, Costagliola2015} emitted from the central star cluster.

Outside the MW, HC$_3$N* has been observed in the nuclei of very active ULIRGs galaxies like  NGC\,4418 \citep{Costagliola2010, Costagliola2015} and Arp\,220 \citep{Martin2011}.
More recently, in \citet{Rico2020} we detected HC$_3$N* emission in the prototypical starburst galaxy NGC\,253.
Like in the MW, the HC$_3$N* emission is related to HCs, but for NGC\,253 their associated luminosities were found to be one order of magnitude brighter than those in the MW and consequently being Super Hot Cores (SHCs). In turn, the presence of SHCs is associated with the very early phases of the formation of Super Star Clusters (i.e. proto Super Star Clusters, proto-SSCs hereafter). SSCs are compact star clusters with ages from $\sim1$ to $100$\,Myr and stellar masses of $\gtrsim10^5$\,M$_\odot$, that represent an extreme mode of star formation. They have been observed mainly in central starbursts of galaxies, Ultra Luminous Infrared Galaxies (ULIRGs) or mergers \citep[see][for a review]{Whitmore1995, Beck2015}.
Despite the 7 proto-SSCs found in  NGC\,253, in \citet{Rico2020} we could not identify any HC$_3$N* emission associated with the strongest non thermal source \citep[TH2,][]{Turner1985} nor its kinematical center \citep{MullerSanchez2010}, both  proposed to host a Super Massive Black Hole (SMBH) in this galaxy.
Unfortunately, the lack of observational evidence of the presence of an active SMBH in NGC\,253 prevented us from studying its possible effects on the vibrational excitation of HC$_3$N.
So far, all the observational data suggest that HC$_3$N* emission is mainly tracing the recent brief episodes of massive star formation in the very early stages of cluster formation. However, there is not any systematic study of the HC$_3$N* emission in nearby galaxies with AGN nuclear activity.

NGC\,1068, one of the closest galaxies hosting spatially resolved AGN and starburst (SB) activities, offers a unique opportunity to study the effects of star formation and AGN activity from its $\sim10^7$\,M$_\odot$ SMBH \citep[][]{Davis2014, Combes2019} on the vibrational excitation of HC$_3$N. NGC\,1068 is a prototypical Seyfert 2 barred galaxy, located at $D=14.4$\,Mpc \citep[][]{BHawthorn1997}, with a luminosity of  $L_\text{IR}=3 \times10^{11}$\,L$_\odot$ \citep{Telesco1980}. Its central region has been extensively observed to study the AGN activity and its effects of the SMBH on its surroundings, including fueling and associated feedback. NGC\,1068 has an AGN driven outflow and bipolar radio jets that strongly interact with the ISM  \citep{SGB2014}. In particular, the elliptical ring with $r\sim200$\,pc, known as the CND, surrounding the SMBH is strongly affected by the AGN driven jets and outflow \citep{Krips2006, SGB2014, SGB2017, SGB2019, Viti2014}.
In addition to the AGN dominated central region, there is a SB pseudo-ring where  most of the recent massive SF of the galaxy concentrates. The SB psuedo-ring is  located at $r\simeq1.3$\,kpc from the nucleus and is formed by a two-armed spiral structure that begins at both ends of the central bar \citep{SGB2014}. The SB pseudo-ring is clearly seen in Pa$\alpha$ and CO, contributing significantly to the total CO luminosity of the galaxy \citep[see Fig.1a from][or Fig~\ref{fig:CO}]{SGB2014}.

\citet{SGB2014} and \citet{Viti2014} analyzed the molecular emission from both the CND and the SB pseudo-ring and found differences in the molecular line ratios (from CO, HCO$^+$, HCN and CS) 
between both regions, indicating that the radiative and mechanical feedback from the AGN has changed  the physical conditions and the chemistry of the molecular gas in the CND. They found that, in general, more dense ($10^5$\,cm$^{-3}$) and warmer ($T\sim150$\,K) gas is found in the CND \citep{Viti2014}, although they only analyze one star-forming region as representative of the SB pseudo-ring. 
Furthermore, the HC$_3$N abundance in the CND has been found to be enhanced likely due to the AGN induced chemistry \citep{Viti2014}, which offers a unique opportunity to study in detail the effects of the SMBH on the heating of its surroundings as traced by the HC$_3$N* emission. 

In this paper we study the HC$_3$N emission in the CND and the SB pseudo-ring of NGC\,1068. Despite  the bright HC$_3$N emission from the ground state observed in the CND, we  find  HC$_3$N vibrationally excited emission in only one condensation of the SB pseudo-ring, revealing the presence of a Super Hot Core (SHC). We show that, contrary to massive proto-stars, the AGN is extremely efficient in heating the gas of the CND mainly through shocks originating in the jet/outflow system, but very inefficient in heating the dust by radiation, which is the requirement for HC$_3$N to be vibrationally excited for the derived H$_2$ densities.

\section{Data reduction}

To study the HC$_3$N emission from the SB pseudo-ring and the CND of NGC\,1068, we have used publicly available data from the ALMA science archive. The observations  are summarized in Table~\ref{tab:observations}. Additionally, we also made use of the HST NICMOS (NIC3) narrow-band (F187N, F190N) Pa\,$\alpha$ line emission image of NGC\,1068 with an angular resolution of $0.26\arcsec\times0.26\arcsec$  \citep[for details on the calibration and imaging, see ][]{SGB2014}.

The ALMA data calibration and imaging  was carried using the  Common Astronomy Software Applications \citep[\texttt{CASA};][]{McMullin2007} 4.2 version pipeline. Continuum emission maps were obtained by averaging line-free channels.  Due to the large amount of molecular emission and the large velocity gradients between the CND and the SB pseudo-ring data were not continuum subtracted in the uv-plane. For deconvolution we have used the \texttt{CASA} \texttt{tclean} task with Briggs weighting setting the \texttt{robust} parameter to $0.5$. All the produced data cubes and continuum maps were corrected for the primary beam.
The achieved synthesized beam sizes and resulting rms noise in the cubes, as well as the corresponding primary beam FWHM (FOV), are also listed in Table~\ref{tab:observations}.
Since most of the NGC\,1068 observations are centered on the AGN position, as we move to higher frequencies the observed field of view of the telescope primary beam is reduced, limiting the observations of the SB pseudo-ring to the outer edge of the mapped area at $\sim220$\,GHz. As a consequence there is a decrease in sensitivity at the edges of the map where the SB pseudo-ring is observed. Fortunately, project-ID 2011.0.00083.S \citep[$\sim350$\,GHz;][]{SGB2014} is an eleven-field mosaic that fully covers the SB pseudo-ring.

For further inspection of the reduced data cubes and spectral line identification and analysis, we have used \texttt{MADCUBA}\footnote{Madrid Data Cube Analysis (\texttt{MADCUBA}) is a software developed in the Center of Astrobiology (Madrid) to visualize and analyze data cubes and single spectra.  Website: \url{https://cab.inta-csic.es/madcuba/}} \citep{Madcuba2019}. Order $0-1$ polynomial baselines were subtracted from the extracted spectra in the regions of interest. This guarantees a flat spectral baseline in order to carry out the molecular line analysis.

\begin{table}
  \caption{ALMA observations used in this work.}
  \centering
	\centering
    \setlength{\tabcolsep}{2pt}
	\label{tab:observations}
	\begin{tabular}{lr @{\ $-$\ } rccc} 
		\hline
		Project-ID  & \multicolumn{2}{c}{Frequency} & Resolution & rms & FOV\\
         &  \multicolumn{2}{c}{(GHz)} & (arcsec) &  (mJy beam$^{-1}$) & (arcsec)\\
		\hline
		2013.1.00055.S		& 85.37 & 100.99	& $0.73\arcsec\times0.46\arcsec$	& 0.67 & 62.49 \\
		2013.1.00221.S  	& 85.55 & 98.54 	& $0.72\arcsec\times0.46\arcsec$	& 0.37 & 63.26 \\
		2013.1.00279.S  	& 88.61 & 104.30 	& $0.67\arcsec\times0.43\arcsec$	& 0.54 & 60.37 \\
        2013.1.00279.S  	& 92.54 & 108.24 	& $0.60\arcsec\times0.38\arcsec$	& 0.96 & 58.00 \\
		2013.1.00060.S  	& 96.25 & 110.05 	& $1.15\arcsec\times0.99\arcsec$	& 0.62 & 56.45 \\
        2013.1.00221.S  	& 128.83 & 130.71 	& $0.45\arcsec\times0.35\arcsec$	& 0.37 & 44.87 \\
        2015.1.01144.S  	& 143.62 & 159.44 	& $0.65\arcsec\times0.44\arcsec$	& 0.62 & 38.43 \\
        2016.1.00232.S  	& 215.35 & 231.97 	& $0.33\arcsec\times0.26\arcsec$	& 0.43 & 26.04 \\
        2011.0.00083.S      & 342.27 & 357.92   & $0.42\arcsec\times0.55\arcsec$    & 3.48 & 48.69$^*$ \\
        2016.1.00232.S  	& 343.56 & 358.29 	& $0.22\arcsec\times0.19\arcsec$	& 0.77 & 16.59 \\
        
		\hline
	\end{tabular}
	\begin{tablenotes}
            \small
            \item \textsuperscript{*} Mosaic.
    \end{tablenotes}
\end{table}

\begin{figure*}
	\includegraphics[width=1\linewidth]{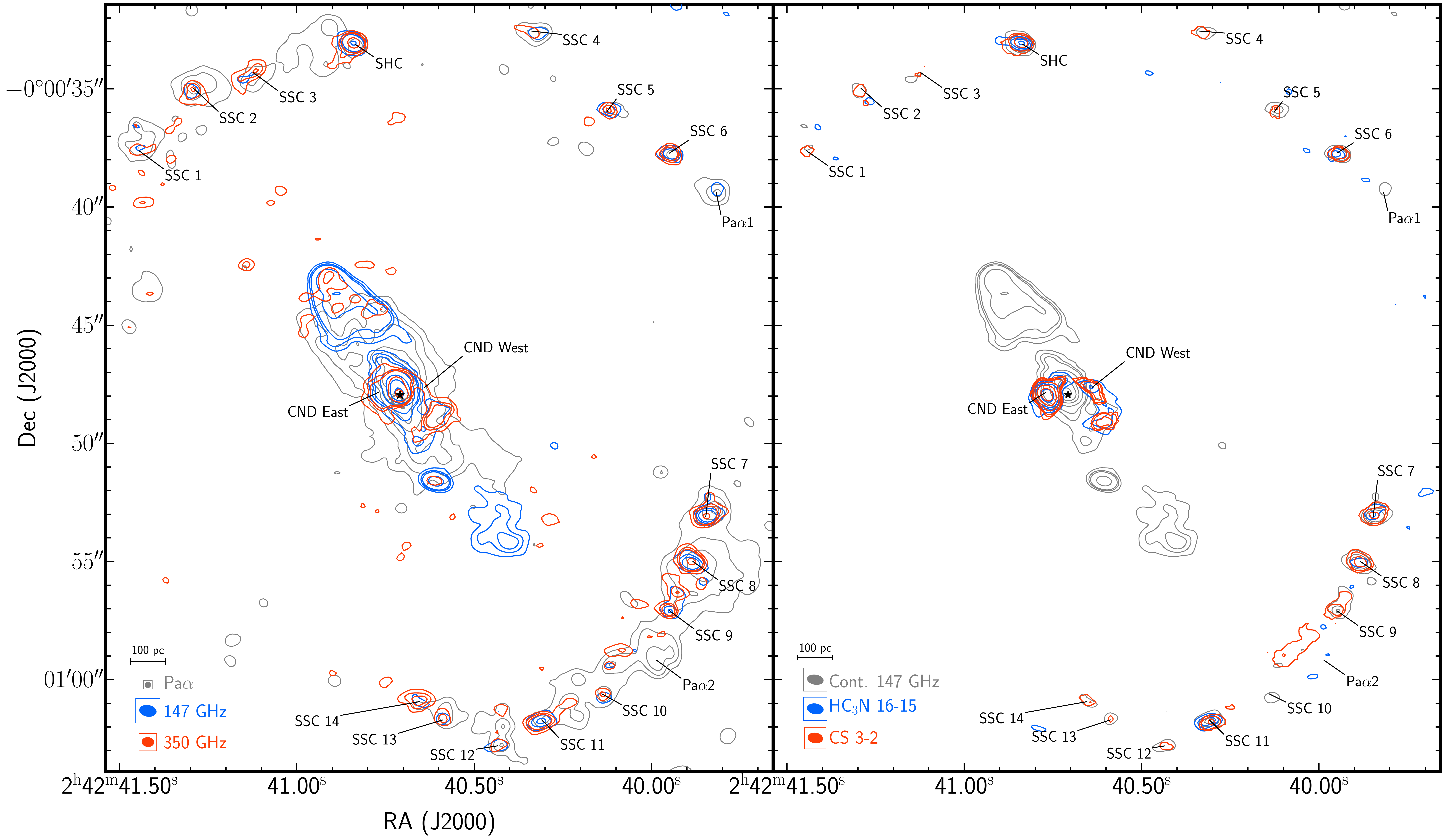}
	\caption{Left panel: (sub)millimiter continuum and Pa\,$\alpha$ emission map of NGC\,1068. The continuum map at $147$\,GHz is shown in blue, with contour levels $5\times$, $10\times$, $15\times$, $50\times$, $150\times$, $300\times$, $600\times \sigma_{147}$, 
	with $\sigma_{147}=0.025$\,mJy\,beam$^{-1}$. The continuum map at $350$\,GHz is shown in red  with contour levels  $5\times$, $10\times$, $15\times$, $50\times$, $150\times \sigma_{350}$, 
	with $\sigma_{350}=0.10$\,mJy\,beam$^{-1}$. 
    Pa\,$\alpha$ emission is shown in grey  with contour levels $10\times$, $20\times$, $40\times$, $80\times$, $160\times$, $320\times \sigma_{\text{Pa}\alpha}$, 
    with $\sigma_{\text{Pa}\alpha}=2\times10^{-16}$\,erg\,cm$^{-2}$\,s$^{-1}$. 
    Right panel: Integrated intensity map of HC$_3$N $v=0$ $(16-15)$ (in blue) and CS $(3-2)$ (in red) line emission over the $147$\,GHz continuum map (in grey).
    The star marks the AGN position. Indicated are also the $14$ continuum clumps identified on the SB pseudo-ring, the Super Hot Core (SHC) position, the $2$ positions with Pa\,$\alpha$ emission but no $350$\,GHz continuum and the $2$ CND positions studied. The beams of each map are shown on the lower left corner of each panel.
    }
    \label{fig:positions}
\end{figure*}

\section{Results}
\subsection{Pa\,$\alpha$ and continuum emission}
\label{subsec:Paschen_Continuum}

The left panel of Figure~\ref{fig:positions} shows the ALMA continuum emission maps obtained at $147$\,GHz and at $350$\,GHz overlaid on the Pa\,$\alpha$ emission.  The figure also labels the main identified features in NGC\,1068 discussed in this paper.
As expected, the Pa\,$\alpha$ emission shows a spatial distribution different from the mm-submm continuum emission. This is partially due to the extinction of the Pa\,$\alpha$ lines by dust, and also because it traces a less embedded and more evolved stage in the process of massive star formation in the SB pseudo-ring (S\'anchez-Garc\'ia et al. in prep.).

From the continuum maps at $147$\,GHz and $350$\,GHz, we have identified $15$ clumps above $5 \sigma$  in the SB pseudo-ring. We have named these positions SSCs because they have stellar masses $\sim10^5$ (Sec.~\ref{sec:stellar_masses}). In addition,  we have detected HC$_3$N* emission in one of these SSCs, which we named SHC  (Sec.~\ref{sub:hc3n_emission}). 

We have also selected two representative sources that are bright in Pa\,$\alpha$ (Pa\,$\alpha$\,1 and Pa\,$\alpha$\,2) but do not have any associated $350$\,GHz  emission. In addition, we have also included two CND positions, located East and West from the AGN. Their coordinates are listed in Table~\ref{tab:table_positions_1}.

To perform the analysis of the continuum emission, we have smoothed the Pa\,$\alpha$ and $350$\,GHz images to the same spatial resolution of the $147$\,GHz map (i.e. $0.62\arcsec\times0.42\arcsec$). We have then measured the peak continuum emission in the three maps at the location of the  $147$\,GHz maximum (see Table~\ref{tab:continuum_fluxes}).

The continuum emission at $147$\,GHz in the CND is mainly dominated by the non-thermal synchrotron emission from the AGN and the bipolar radio jets interacting with the ISM and its AGN-driven outflow \citep{SGB2014}.
In the SB pseudo-ring, the continuum emission at $147$\,GHz is expected to be mainly dominated by free-free emission.  The flux at $8.4$\,GHz measured with the VLA with an angular resolution of $3.5\arcsec\times2.9\arcsec$ at the SHC is $\sim1.6$\,Jy \citep{Anantharamaiah1993}, which is consistent with optically thin free-free emission at $147$\,GHz as the spectral index between these frequencies is $\sim-0.1$. However, the beam sizes are rather different. We can also use the continuum emission at $350$\,GHz to estimate the possible contribution from dust emission to the $147$\,GHz continuum emission. Assuming that the dust emission measured at $350$\,GHz is optically thin and a typical dust emissivity spectral index of $1.5$, the expected contribution of the dust emission to the $147$\,GHz flux will be only of about $20\%$, i.e the $147$\,GHz continuum is dominated by the free-free emission.  In the following discussion, we  will consider that the continuum emission at $350$\,GHz is dominated by dust emission both toward the CND and the SB pseudo-ring.
Both continuum emissions coincide spatially on the SB pseudo-ring, where they exhibit a clumpy structure associated to star-forming regions. Note that non-thermal emission dominates throughout the radio jet trajectory \citep[which  encloses  the CND East knot; for a detailed analysis on the CND spectral indexes see ][]{SGB2019}.

\begin{table}
    \begin{center}
    \caption[]{Coordinates and velocities of the analyzed clumps.} 
    \label{tab:table_positions_1}
    \setlength{\tabcolsep}{4pt}
    \begin{tabular}{lccc}
        \hline \noalign {\smallskip}
Location	&	RA	        &	Dec	            &	V$_\text{LSR}$		\\
            &   (J2000)             &   (J2000)         &     (km\,s$^{-1}$)    \\
            &   $02^{\text{h}}$\,$42^{\text{m}}$            &   $-00^{\circ}$       &     \\
        \hline \noalign {\smallskip}
SHC     	&	$40^{\text{s}}.84$	&	$00\arcmin$\,$32\arcsec.94$	& $	1094.3	\pm	0.3	$ \\
SSC 1	    &	$41^{\text{s}}.44$	&	$00\arcmin$\,$37\arcsec.48$	& $	1004.4	\pm	1.2	$ \\
SSC 2	    &	$41^{\text{s}}.29$	&	$00\arcmin$\,$34\arcsec.97$	& $	1054.8	\pm	1.1	$ \\
SSC 3	    &	$41^{\text{s}}.15$	&	$00\arcmin$\,$34\arcsec.64$	& $	1067.0	\pm	2.3	$ \\
SSC 4	    &	$40^{\text{s}}.32$	&	$00\arcmin$\,$32\arcsec.66$	& $	1168.0	\pm	0.9	$ \\
SSC 5	    &	$40^{\text{s}}.12$	&	$00\arcmin$\,$35\arcsec.87$	& $	1174.1	\pm	1.1	$ \\
SSC 6	    &	$39^{\text{s}}.95$	&	$00\arcmin$\,$37\arcsec.75$	& $	1193.2	\pm	0.6	$ \\
SSC 7 	&	$39^{\text{s}}.84$	&	$00\arcmin$\,$53\arcsec.12$	& $	1287.1	\pm	0.3	$ \\
SSC 8	    &	$39^{\text{s}}.88$	&	$00\arcmin$\,$54\arcsec.96$	& $	1277.3	\pm	0.4	$ \\
SSC 9	    &	$39^{\text{s}}.95$	&	$00\arcmin$\,$57\arcsec.88$	& $	1263.5	\pm	0.3	$ \\
SSC 10	&	$40^{\text{s}}.13$	&	$01\arcmin$\,$00\arcsec.76$	& $	1250.2	\pm	1.0	$ \\
SSC 11	&	$40^{\text{s}}.30$	&	$01\arcmin$\,$01\arcsec.64$	& $	1193.0	\pm	0.5	$ \\
SSC 12	&	$40^{\text{s}}.44$	&	$01\arcmin$\,$02\arcsec.85$	& $	1143.9	\pm	0.7	$ \\
SSC 13	&	$40^{\text{s}}.59$	&	$01\arcmin$\,$01\arcsec.61$	& $	1135.1	\pm	1.6	$ \\
SSC 14	&	$40^{\text{s}}.65$	&	$01\arcmin$\,$00\arcsec.95$	& $	1126.4	\pm	1.5	$ \\
Pa\,$\alpha$ 1	&	$39^{\text{s}}.81$	&	$00\arcmin$\,$39\arcsec.37$	& $	1205.8	\pm	0.8	$  \\
Pa\,$\alpha$ 2	&	$39^{\text{s}}.97$	&	$00\arcmin$\,$59\arcsec.28$	& $	1247.3	\pm	0.9	$ \\
CND E.	&	$40^{\text{s}}.77$	&	$00\arcmin$\,$47\arcsec.84$	& $	1076.3	\pm	0.6	$ \\
CND W.	&	$40^{\text{s}}.64$	&	$00\arcmin$\,$47\arcsec.64$	& $	1191.8	\pm	1.1	$ \\
        \hline \noalign {\smallskip}
    \end{tabular}
        \begin{tablenotes}
            \small
            \item \textsuperscript{a} Velocity derived from CS since no HC$_3$N was detected.
        \end{tablenotes}
    \end{center}
\end{table}

\begin{table}
    \begin{center}
    \caption[]{Peak flux densities of the Pa$\alpha$ and $147$\,GHz/$350$\,GHz continuum emission measured toward the position of the $147$\,GHz continuum peak for each source. The three emission maps have been smoothed to a common resolution of $0.62\arcsec\times0.42\arcsec$ in order to measure the fluxes. Pa$\alpha$ is measured in $10^{-14}$\,erg\,cm$^{-2}$\,s$^{-1}$, while Cont. $147$ and Cont. $350$ in mJy\,beam$^{-1}$. $\theta^2_{350}$, in arcsec$^2$ ($1\arcsec \sim 70$\,pc), is the deconvolved surface area obtained from fitting a two-dimensional Gaussian to the $350$\,GHz continuum emission.}
    \label{tab:continuum_fluxes}
    \setlength{\tabcolsep}{3pt}
    \begin{tabular}{lccccc}
    
        \hline \noalign {\smallskip}
        Location	&	Pa$\alpha$	&	Cont. $147$	&	Cont. $350$	& 	$\theta^2_{350}$\\
        
        \hline \noalign {\smallskip}
        SHC	     & $1.42 \pm 0.01$ & $1.37 \pm 0.03 $ & $ 7.54 \pm 0.12 $ & $	0.26	\pm	0.04 $ \\
        SSC 1  & $1.96 \pm 0.02$ & $0.14 \pm 0.02 $ & $ 0.65 \pm 0.12 $ & $  0.54    \pm 0.31 $ \\
        SSC 2	 & $3.17 \pm 0.02$ & $0.21 \pm 0.02 $ & $ 1.46 \pm 0.12 $ & $	0.72	\pm	0.26 $ \\
        SSC 3  & $0.83 \pm 0.02$ & $0.16 \pm 0.02 $ & $ 1.20 \pm 0.12 $ & $  1.03    \pm 0.22 $ \\
        SSC 4  & $1.93 \pm 0.02$ & $0.25 \pm 0.03 $ & $ 0.65 \pm 0.09 $ & $  0.77    \pm 0.28 $ \\
        SSC 5  & $0.71 \pm 0.03$ & $0.35 \pm 0.03 $ & $ 1.47 \pm 0.09 $ & $  0.23    \pm 0.08 $ \\
        SSC 6	 & $0.71 \pm 0.02$ & $0.72 \pm 0.03 $ & $ 3.18 \pm 0.09 $ & $	0.26	\pm	0.04 $ \\
        SSC 7	 & $3.50 \pm 0.02$ & $0.76 \pm 0.03 $ & $ 5.34 \pm 0.12 $ & $	0.36	\pm	0.05 $ \\
        SSC 8	 & $3.31 \pm 0.02$ & $0.72 \pm 0.03 $ & $ 5.78 \pm 0.13 $ & $	0.45	\pm	0.18 $ \\
        SSC 9	 & $1.03 \pm 0.02$ & $0.39 \pm 0.03 $ & $ 2.51 \pm 0.10 $ & $	0.16	\pm	0.09 $ \\
        SSC 10 & $1.60 \pm 0.01$ & $0.25 \pm 0.03 $ & $ 0.91 \pm 0.11 $ & $  0.23    \pm 0.10 $ \\
        SSC 11 & $1.21 \pm 0.01$ & $0.42 \pm 0.03 $ & $ 3.28 \pm 0.11 $ & $	0.50	\pm	0.12 $ \\
        SSC 12 & $1.27 \pm 0.01$ & $0.21 \pm 0.03 $ & $ 0.67 \pm 0.11 $ & $  0.57    \pm 0.20 $ \\
        SSC 13 & $1.71 \pm 0.02$ & $0.19 \pm 0.03 $ & $ 1.22 \pm 0.11 $ & $  0.33    \pm 0.10 $ \\
        SSC 14 & $0.64 \pm 0.02$ & $0.18 \pm 0.03 $ & $ 1.94 \pm 0.11 $ & $  0.79    \pm 0.13 $ \\
        Pa$\alpha$ 1 & $2.19 \pm 0.03$ & $0.18 \pm 0.04 $ & $ \leqslant 0.26 $  & $	0.43	\pm	0.38	$ \\
        Pa$\alpha$ 2	 & $	1.37 \pm 0.02	$ & $ \leqslant	0.13	$ & $ \leqslant	0.35	$ & $ 	0.29	\pm	0.27	$ \\
        CND E.	& $6.57 \pm 0.01$ & $0.65 \pm 0.01 $ & $ 3.72 \pm 0.08 $  & - \\
        CND W.	& $5.31 \pm 0.03$ & $0.12 \pm 0.01 $ & $ 0.51 \pm 0.08 $  & - \\
        \hline \noalign {\smallskip}
    \end{tabular}
    \end{center}
\end{table}

\subsection{HC$_3$N emission}
\label{sub:hc3n_emission}

The right panel of Figure~\ref{fig:positions} shows the distribution of the HC$_3$N $v=0$ $(J=16-15)$ and the CS $(J=3-2)$ velocity-integrated line emission overlaid on the $147$\,GHz continuum  emission. This figure shows that the HC$_3$N $v=0$ ($J=16-15$) and CS ($J=3-2$)  lines trace the CND and the high density star formation clumps in the SB pseudo-ring and closely follows the $147$\,GHz continuum in those regions. 
HC$_3$N emission is detected in sources with radio continuum and/or Pa\,$\alpha$ emission, namely positions CND E., CND W., SHC and SSC 6, 7, 8, 9 and 11.
We have measured the velocity of the gas for the positions in Table~\ref{tab:table_positions_1} from HC$_3$N $v=0$ emission (when detected) and from CS. Other molecules such as CO, HCN, HNCO, H$_2$CO or CH$_3$OH  were also detected, but in this paper we will focus on the HC$_3$N emission of the selected positions.

\begin{table}
    \begin{center}
    \caption[]{HC$_3$N* transitions quantum numbers and frequencies present in the analyzed data.} 
    \label{tab:hc3n_quants}
    \setlength{\tabcolsep}{4pt}
    \begin{tabular}{lcccrrc}
        \hline \noalign {\smallskip}
Vib. State	&	$J_\text{up}-J_\text{low}$	&	$l$-doubling	&	Parity	&	Frequency	&	$E_\text{up}$	&	$g_\text{up}$	\\
	&		&	$(l_\text{up}-l_\text{low})$	&		&	(GHz)	&	(K)	&		\\
	 \hline \noalign {\smallskip}
$v=0$ 	&	 10-9 	&		&		&	90.98	&	24	&	21	\\
$v=0$ 	&	 11-10 	&		&		&	100.08	&	29	&	23	\\
$v=0$ 	&	 12-11 	&		&		&	109.17	&	34	&	25	\\
$v=0$ 	&	 16-15 	&		&		&	145.56	&	59	&	33	\\
$v_6=1$	&	 16-15 	&	 (-1, +1) 	&	1e	&	145.80	&	777	&	33	\\
$v_6=1$	&	 16-15 	&	 (+1, -1) 	&	1f	&	145.91	&	777	&	33	\\
$v_7=1$	&	 16-15 	&	 (-1, +1) 	&	1e	&	145.92	&	380	&	33	\\
$v_7=1$	&	 16-15 	&	 (+1, -1) 	&	1f	&	146.13	&	380	&	33	\\
$v_7=2$	&	 16-15 	&	(0, 0)	&	0	&	146.48	&	701	&	33	\\
$v_7=2$	&	 16-15 	&	(-2, +2) 	&	2e	&	146.48	&	705	&	33	\\
$v_7=2$	&	 16-15 	&	(+2, -2) 	&	2f	&	146.49	&	705	&	33	\\
$v=0$ 	&	 24-23 	&		&		&	218.32	&	131	&	49	\\
$v_6=1$	&	 24-23 	&	 (-1, +1) 	&	1e	&	218.68	&	849	&	49	\\
$v_6=1$	&	 24-23 	&	 (+1, -1) 	&	1f	&	218.85	&	849	&	49	\\
$v_7=1$	&	 24-23 	&	 (-1, +1) 	&	1e	&	218.86	&	452	&	49	\\
$v_7=1$	&	 24-23 	&	 (+1, -1) 	&	1f	&	219.17	&	452	&	49	\\
$v_7=2$	&	 24-23 	&	 (0, 0) 	&	0	&	219.68	&	774	&	49	\\
$v_7=2$	&	 24-23 	&	 (-2, +2) 	&	2e	&	219.71	&	777	&	49	\\
$v_7=2$	&	 24-23 	&	 (+2, -2) 	&	2f	&	219.74	&	777	&	49	\\
$v=0$ 	&	 38-37 	&		&		&	345.61	&	324	&	77	\\
$v=0$ 	&	 39-38 	&		&		&	354.70	&	341	&	79	\\
$v_6=1$	&	 39-38 	&	 (+1, -1) 	&	1e	&	355.28	&	1059	&	79	\\
$v_6=1$	&	 39-38 	&	 (-1, +1) 	&	1f	&	355.56	&	1059	&	79	\\
$v_7=1$	&	 39-38 	&	 (+1, -1) 	&	1e	&	355.57	&	662	&	79	\\
$v_7=1$	&	 39-38 	&	 (-1, +1) 	&	1f	&	356.07	&	663	&	79	\\
$v_7=2$	&	 39-38 	&	 (+2, -2) 	&	2e	&	356.94	&	988	&	79	\\
$v_7=2$	&	 39-38 	&	 (-2, +2) 	&	2f	&	357.08	&	988	&	79	\\
        \hline \noalign {\smallskip}
    \end{tabular}
    \end{center}
\end{table}

\begin{figure*}
	\includegraphics[width=0.9\linewidth]{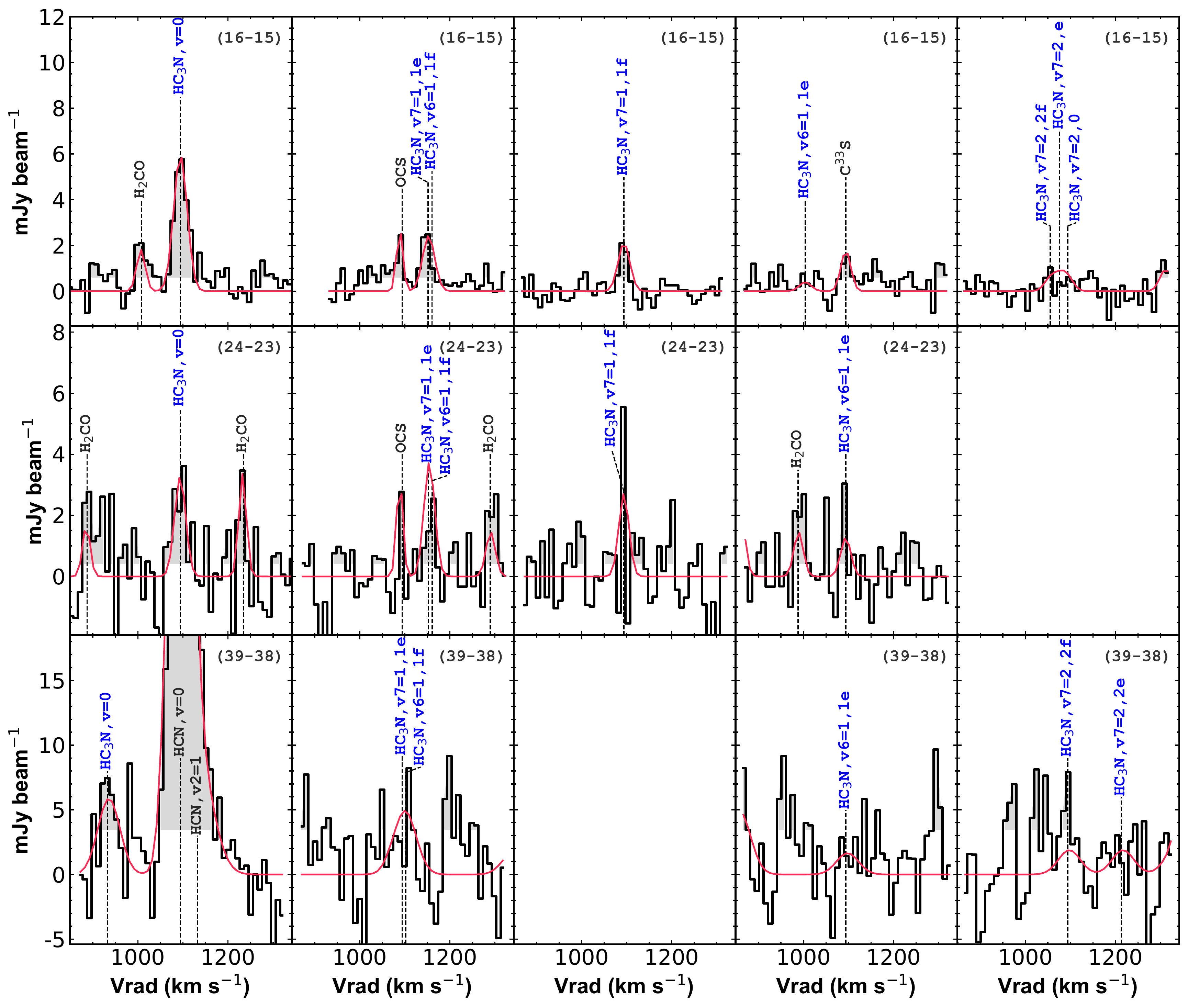}
    \caption{Observed spectra (black histograms) towards the SHC. HC$_3$N* lines are highlighted in blue. Grey shaded spectra represents values above the $1\sigma$ level. Each row corresponds to a different $(J_\text{up}-J_\text{low})$ transition (indicated on the top-right corner of each panel), The top row corresponds to $J=16-15$, the middle row to $J=24-23$ and the bottom row to $J=39-38$.
    HC$_3$N transitions from the same vibrational state and wave function parity but different $(J_\text{up}-J_\text{low})$, are organised in columns.
    Empty panels reflect that there is no data available for that HC$_3$N transition (i.e. the $J=39-38$ HC$_3$N, $v_7=1,1f$ and all $J=24-23$ HC$_3$N, $v_7=2$ transitions). The red lines represent the fitted LTE model obtained from \texttt{MADCUBA}. 
    }
    \label{fig:SHC_spectra}
\end{figure*}

\begin{figure}
	\includegraphics[width=0.9\linewidth]{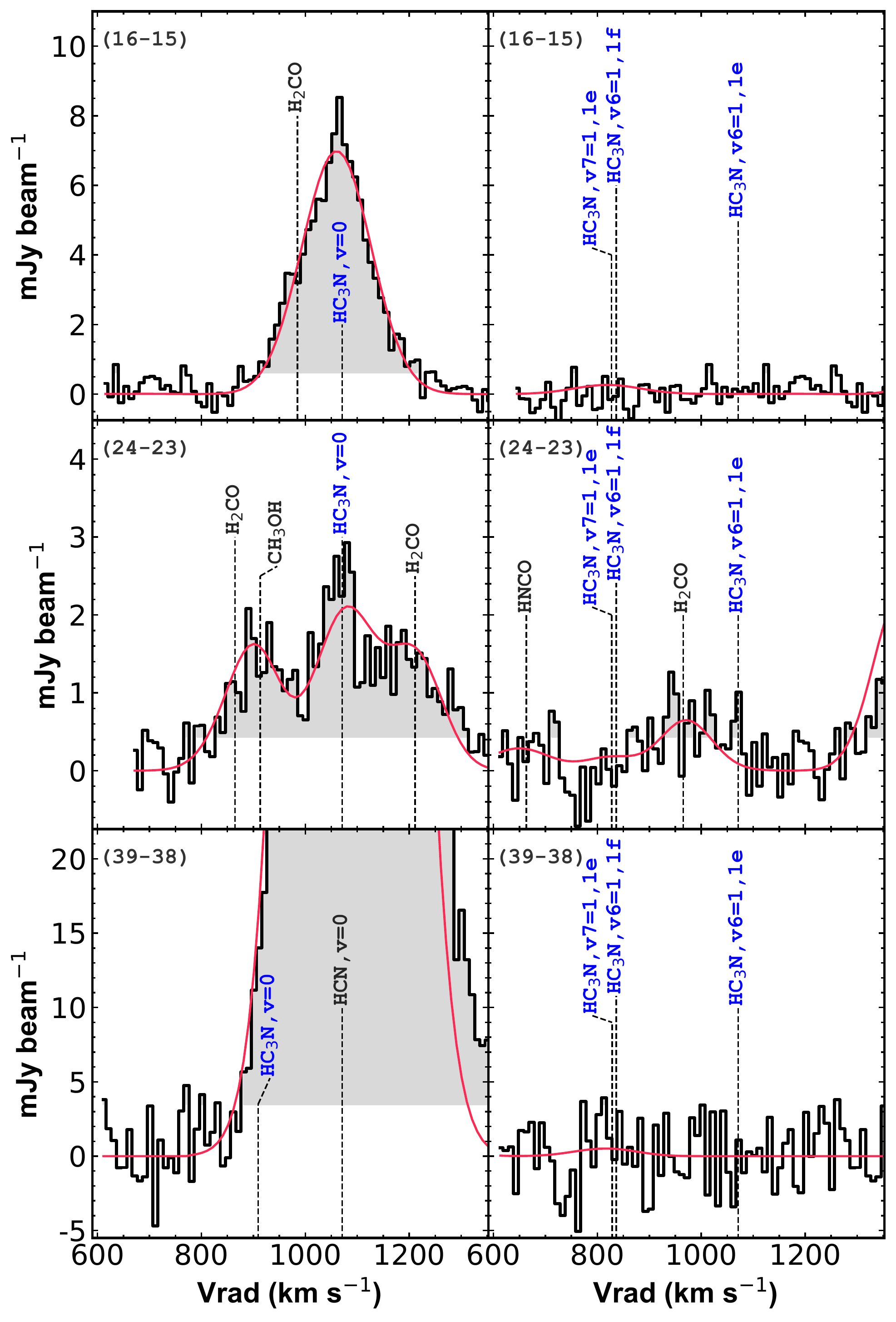}
    \caption{Observed spectra (black histograms) together with the fitted \texttt{MADCUBA} LTE profiles (solid red lines) towards the CND East position. The position of the HC$_3$N transitions are highlighted in blue and blended lines from other identified molecules are shown with grey labels. Grey shaded areas under the profiles represent intensities above the spectra rms.
    Each row corresponds to a different $(J_\text{up}- J_\text{low})$ HC$_3$N transition,  shown in the upper-left corner of every panel.
    The left column corresponds to transitions from the $v=0$ ground state and the right column to the (undetected) transitions from the $v_7=1$ and $v_6=1$ vibrationally excited states.
    }
    \label{fig:CNDe_spectra}
\end{figure}

In addition to the HC$_3$N emission from the ground vibrational state we have searched for HC$_3$N* emission from the SB pseudo-ring and the CND. Table~\ref{tab:hc3n_quants} lists the spectroscopic parameteres the HC$_3$N* transitions observed in our data. We have detected HC$_3$N* emission from the $v_7=1$ vibrationally excited state  in the rotational transitions $J=16-15$ ($E_u=380$\,K) and $J=24-23$ ($E_u=452$\,K) towards only one condensation in the SB pseudo-ring, hereafter the Super Hot Core (SHC). Integrated intensities in the SHC of rotational transitions from the $v_6=1$ vibrationally excited state are just below $1\sigma$ and we treat them as undetected. Figure~\ref{fig:SHC_spectra} displays the spectra of  HC$_3$N  from the rotational transitions $J=16-15$, $J=24-23$ and $J=39-38$ in the ground, $v_7=1$  and  $v_6=1$ vibrational states towards the SHC. The lack of sensitivity prevented us from detecting HC$_3$N* emission towards any other position in the SB pseudo-ring. It is remarkable that, in spite of the enhanced column density of HC$_3$N in the CND (see Sec.~\ref{subsec:derivedHC3Nprops}), we have not detected any HC$_3$N* emission towards the CND positions.  Like Figure~\ref{fig:SHC_spectra},  Figure~\ref{fig:CNDe_spectra} shows the spectra of the rotational transitions in the $v=0$ and $v_7=1$ vibrational states towards CND East. 
The integrated fluxes for the observed HC$_3$N* transitions for the most relevant positions (SHC, SSC 7, CND East and CND West) are listed in Table~\ref{tab:hc3n_fluxes}, with non-detected lines represented as upper limits with their corresponding $3\sigma$ integrated intensity values. It is worth noting the large contrast between the $v=0$ and $v_7=1$ rotational lines.

\begin{table*}
    \begin{center}
    \caption[]{\texttt{MADCUBA} fitted integrated intensity for  HC$_3$N emission lines in mJy\,beam$^{-1}$\,km\,s$^{-1}$ for the SHC, SSC 7, CND East and CND West positions. Upper limits are $3\sigma$.}
    \label{tab:hc3n_fluxes}
    \begin{threeparttable}
    \setlength{\tabcolsep}{3pt}
    \begin{tabular}{l@{\ }lr@{\hspace{0.1\tabcolsep}}lcr@{\hspace{0.1\tabcolsep}}lcr@{\hspace{0.1\tabcolsep}}lcr@{\hspace{0.1\tabcolsep}}lcr@{\hspace{0.1\tabcolsep}}lcr@{\hspace{0.1\tabcolsep}}lcr@{\hspace{0.1\tabcolsep}}lcr@{\hspace{0.1\tabcolsep}}l}
    \hline \noalign {\smallskip}
\multicolumn{2}{c}{Transition}		& \multicolumn{2}{c}{SHC}   &        & 
\multicolumn{2}{c}{SSC 6} &   & \multicolumn{2}{c}{SSC 7} &   & 
\multicolumn{2}{c}{SSC 8} &   & 
\multicolumn{2}{c}{SSC 9} &   & 
\multicolumn{2}{c}{SSC 11} &   & 
\multicolumn{2}{c}{CND E} 		& &	\multicolumn{2}{c}{CND W} 	  \\ 
\hline \noalign {\smallskip}

$v=0$ 	&	10-9	    & $	41	\pm $&$	10	$ & & $ \leqslant$ & $	29	$ & &$	27	\pm  $ & $	6	$ & &$	46	\pm  $ & $	12	$ & &$	44	\pm  $ & $	12	$ & &$	\leqslant$ & $	28	$ & &	-	        &		  & &	-           &	 \\
$v=0$ 	&	11-10	    & $	109	\pm $&$	27	$ & & $	73	\pm  $ & $	15	$ & &$	79	\pm  $ & $	12	$ & &$	89	\pm  $ & $	14	$ & &$	39	\pm  $ & $	12	$ & &$	50	\pm  $ & $	14	$ & &$	477	\pm  $  &$	119	$ & & $	108	\pm   $ & $	27	$\\
$v=0$ 	& 	12-11	    & $	137	\pm $&$	34	$ & & $	60	\pm  $ & $	15	$ & &$	64	\pm  $ & $	16	$ & &$	64	\pm  $ & $	14	$ & &$	56	\pm  $ & $	13	$ & &$	79	\pm  $ & $	15	$ & &$	734	\pm  $  &$	183	$ & & $	308	\pm   $ & $	77	$\\
$v=0$ 	&	16-15	    & $	224	\pm $&$	56	$ & & $	84	\pm  $ & $	11	$ & &$	98	\pm  $ & $	24	$ & &$	73	\pm  $ & $	12	$ & &$	35	\pm  $ & $	12	$ & &$	97	\pm  $ & $	11	$ & &$ 1124 \pm  $  &$	281	$ & & $	306	\pm   $ & $	76	$\\
$v=0$ 	&	24-23	    & $	113	\pm $&$	28	$ & & $	\leqslant$ & $	43	$ & &$	65	\pm  $ & $	16	$ & &$	\leqslant$ & $	47	$ & &$	\leqslant$ & $	43	$ & &$	\leqslant$ & $	46	$ & &$	234	\pm  $  &$	 59	$ & & $  \leqslant$ & $	89			$\\
$v=0$ 	&  	38-37	    & $	381	\pm $&$	95	$ & & $	\leqslant$ & $	184	$ & &$ 	\leqslant$ & $	195	$ & &$ 	\leqslant$ & $	191	$ & &$ 	\leqslant$ & $	172	$ & &$ 	\leqslant$ & $	180	$ & &$ 	\leqslant$  &$ 1769	$\textsuperscript{a} & & $ \leqslant$ & $	243			$\textsuperscript{a}  \\
$v=0$  	&  	39-38	    & $	282	\pm $&$	70	$ & & $	\leqslant$ & $	171	$ & &$ 	\leqslant$ & $	181	$ & &$ 	\leqslant$ & $	178	$ & &$ 	\leqslant$ & $	160	$ & &$ 	\leqslant$ & $	168	$ & &$ 	\leqslant$  &$	304	$\textsuperscript{b} & & $ \leqslant$ & $	350			$\textsuperscript{b}\\
$v_7=1$	& 	16-15 1e	& $	59	\pm $&$	15	$ & & $	\leqslant$ & $	32	$ & &$ 	\leqslant$ & $	33	$ & &$ 	\leqslant$ & $	35	$ & &$ 	\leqslant$ & $	31	$ & &$ 	\leqslant$ & $	33	$ & &$ 	\leqslant$  &$	 82 $ & & $ \leqslant$ & $	86			$\\
$v_7=1$ &	16-15 1f	& $	58	\pm $&$	15	$ & & $	\leqslant$ & $	32	$ & &$ 	\leqslant$ & $	33	$ & &$ 	\leqslant$ & $	35	$ & &$ 	\leqslant$ & $	31	$ & &$ 	\leqslant$ & $	33	$ & &$ 	\leqslant$  &$	 82	$ & & $ \leqslant$ & $	86			$\\
$v_7=1$	&	24-23 1e	& $	50	\pm $&$	13	$ & & $	\leqslant$ & $	42	$ & &$ 	\leqslant$ & $	46	$ & &$ 	\leqslant$ & $	46	$ & &$ 	\leqslant$ & $	41	$ & &$ 	\leqslant$ & $	44	$ & &$ 	\leqslant$  &$	 81	$ & & $ \leqslant$ & $	89			$\\
$v_7=1$	&	24-23 1f	& $	50	\pm $&$	13	$ & & $	\leqslant$ & $	42	$ & &$  \leqslant$ & $	46	$ & &$ 	\leqslant$ & $	46	$ & &$ 	\leqslant$ & $	41	$ & &$ 	\leqslant$ & $	44	$ & &$ 	\leqslant$  &$	 81	$ & & $ \leqslant$ & $	89			$\\

        \hline \noalign {\smallskip}
    \end{tabular}
    \begin{tablenotes}
      \small
      \item \textsuperscript{a} Strongly blended with CO(3-2) due to the higher FWHMs in the CND (see Table~\ref{tab:table_hc3n_lums}).
      \item \textsuperscript{b} Blended with HCN(4-3) due to the higher FWHMs in the CND (see Table~\ref{tab:table_hc3n_lums}).
    \end{tablenotes}
    \end{threeparttable}
    \end{center}
\end{table*}

\section{Analysis}

\subsection{Stellar and gas masses in the SB pseudo-ring}
\label{sec:stellar_masses}

We can use the Pa\,$\alpha$ and the continuum emission at $147$\,GHz dominated by free-free emission to estimate the stellar mass of massive stars arriving the main sequence. Using the canonical values for an \ion{H}{ii} region of $T_e=10^4$\,K and electron density $n_e=10^4$\,cm$^{-3}$, the production rate of ionizing photons from the Pa\,$\alpha$ peak emission, $Q^{0}_{L_{\text{Pa}\alpha}}$ is \citep{Kennicutt1998, Osterbrock2006}:
\begin{align}
Q^{0}_{L_{\text{Pa}\alpha}}(\text{s}^{-1}) = 7.344\times10^{12} \cdot \left(\frac{L_{\text{Pa}\alpha}}{\text{erg} \; \text{s}^{-1}}\right)
\end{align}
The values can change $\sim15\%$ due to variations on the electron temperature within $T_e=5000-20000$\,K  and remain insensitive for electron density variations within $n_e=10^2-10^6$\,cm$^{-3}$ \citep{Osterbrock2006, Piqueras2016}.
We can also obtain the production rate of ionizing photons from the continuum emission at $147$\,GHz, $Q^{0}_{147}$, using the approach by \citet{Murphy2011} and assuming the same $T_e$ as for the determination of the $Q^{0}_{L_{\text{Pa}\alpha}}$:
\begin{align}
    Q^{0}_{147}(\text{s}^{-1}) = 10^{26} \left(\frac{T_e}{10^4\:\text{K}}\right)^{-0.45} \left(\frac{\nu}{147\text{GHz}}\right)^{0.1} \times \left(\frac{L_{147}}{\text{erg}\;\text{s}^{-1}\;\text{Hz}^{-1}}\right)
\end{align}
Finally, we convert the ionizing photons production rates to Zero Age Main Sequence (ZAMS) stellar masses ($M_*$) following \citet{Leroy2018}:
\begin{align}
M_* (\text{M}_\odot) \sim \frac{Q^0}{4\times10^{46}}
\end{align}

The ZAMS stellar masses derived from the ionizing photons production rates  from Pa\,$\alpha$ and $147$\,GHz continuum emission ($M_{*,\text{Pa}\alpha}$ and $M_{*,147}$), are compared on Table~\ref{tab:summary_masses}. The mass of ZAMS stars in all condensations with radio continuum emission are a few $10^5$\,M$_\odot$, typical of the Super Star Clusters (hereafter SSCs, young star clusters with stellar masses $\gtrsim 10^4$\,M$_\odot$) found in NGC\,253 \citep{Rico2020}, indicating that a substantial fraction of the star formation in the SB pseudo-ring seems to be dominated by SSCs. 

On the other hand, we can use the $350$\,GHz continuum emission to estimate the total gas mass from dust emission. Following \citet{Leroy2018}, we first estimate the optical depth ($\tau_{350}$) by comparing the measured intensity peak ($I_{350}$) with that expected by assuming a dust temperature: 
\begin{align}
I_{350}= (1-e^{\tau_{350}})\:B_\nu(T_\text{dust})
\end{align}
where $B_\nu$ is the blackbody intensity at $350$\,GHz for a given dust temperature. For our mass estimates we have assumed a dust temperature of $80$\,K \citep[see][for SSCs in NGC\,253]{Leroy2018}. Changes in a factor of $2$ in $T_\text{dust}$ will change the masses by a similar factor. Assuming a mass absorption coefficient $\kappa_{350}=1.9$\,cm$^{2}$\,g$^{-1}$ \citep[][]{Leroy2018} and a standard dust-to-gas ratio (DGR) of $1/100$ by mass (close to the value found by \citet{Wilson2008} for SB galaxies and similar to the $1/150$ value commonly used for the Milky Way), we can convert $\tau_{350}$ into a gas surface density ($\Sigma_\text{gas}$) which is then converted  into gas mass from the cloud surface area ($\theta^2$):
\begin{align}
    M_\text{gas} (\text{M}_\odot) =\theta^2 \Sigma_\text{gas}=\theta^2 \frac{1}{\text{DGR}\cdot \kappa_{350}}\tau_{350}
\end{align}
The cloud surface areas have been derived from a two-dimensional Gaussian fitting to the  $350$\,GHz continuum emission after deconvolution of the beam profile.
The values of $\theta
^2$ are listed in Table~\ref{tab:continuum_fluxes} and the gas masses in  Table~\ref{tab:summary_masses}.

\subsection{Physical properties derived from HC$_3$N}
\label{subsec:derivedHC3Nprops}

Following the procedure used for NGC\,253 in \citet{Rico2020}, we have used both LTE and non-LTE multiline analysis of the HC$_3$N emission from the $v=0$ and the $v_7=1$ vibrational states to derive the physical properties of the star-forming regions in the SB pseudo-ring and the CND. We analyze the HC$_3$N emission in sources where emission from the $v=0$ is detected and we also include some positions with no HC$_3$N emission for completeness (e.g. SSC\,2, Pa$\alpha$\,1 and Pa$\alpha$\,2). 

The excitation of HC$_3$N can be dominated by different mechanisms: while its vibrational transitions are pumped mainly via absorption of mid-IR photons (IR pumping), the excitation of the rotational levels within a given vibrational state is usually dominated by collisions with H$_2$ \citep{Rico2020}.

\subsubsection{LTE analysis}

To describe both types of excitation two different excitation temperatures are used, in LTE the vibrational temperature ($T_\text{vib}$) describing the excitation between vibrational levels, and the rotational temperature ($T_\text{rot}$) describing the excitation between rotational levels within a vibrationally excited state. Usually, $T_\text{vib}$ reflects the dust temperature in the case of IR pumping \citep{Rico2020}.

\begin{figure}
\centering
    \includegraphics[width=\linewidth]{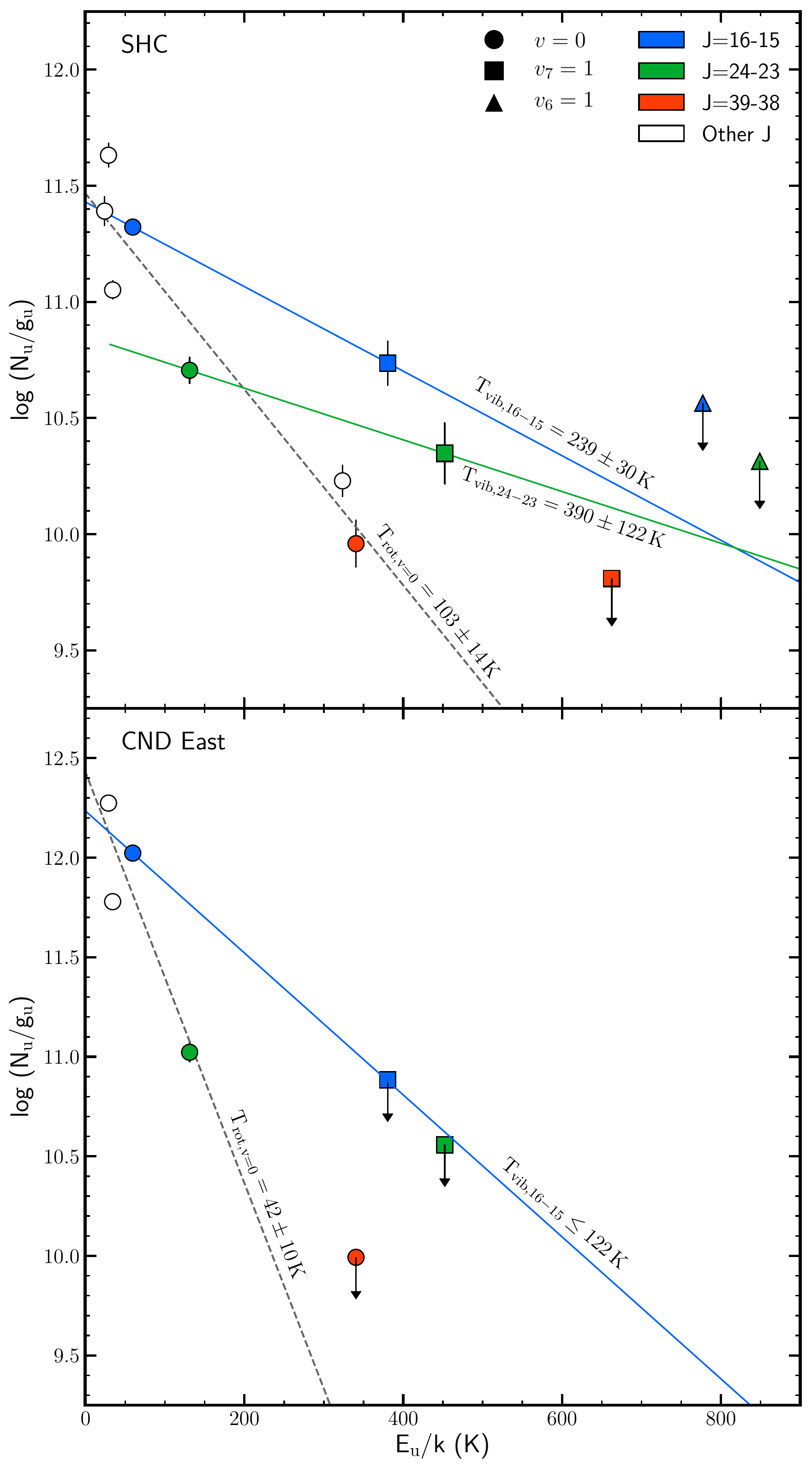}
  \caption{Rotational diagram derived from the line intensities of HC$_3$N$^*$ for the SHC position (upper panel) and CND East position (lower panel). 
  Transitions from the ground state $v=0$ are marked with circles, $v_7=1$ with squares and  $v_6=1$ with triangles. The  $J = 39-38$ transitions are highlighted in red, $J = 24-23$ in green and $J = 16-15$ in blue. Dashed grey lines represent the fit to all rotational transitions $(J_\text{up}-J_\text{low})$ from the ground state $v=0$ (i.e. $T_{\text{rot},v=0}$). The blue and green solid line represents the fit to the $J = 16-15$ and the $J = 24-23$ transitions from the $v=0$ and $v_7=1$ states, respectively (i.e. $T_{\text{vib},16-15}$ and $T_{\text{vib},24-23}$). Errors have been obtained from bootstrapping.
  }
  \label{fig:rotdiag}
\end{figure}

The need of two excitation temperatures to describe the excitation of HC$_3$N is illustrated in the rotational diagram (Figure~\ref{fig:rotdiag}). The rotational diagram can be used to obtain total column densities and excitation temperatures from line integrated intensities. A detailed explanation on the rotational diagram method can be found in \citet{Goldsmith1999}. In short, assuming that the emission is optically thin, HC$_3$N column densities from the upper level can be derived as:
\begin{align}
    N_u^\text{thin}=\frac{8\pi k \nu^2 W}{h c^3 A_\text{ul}}
\end{align}
where $W$ is the integrated line intensity (in K\,km\,s$^{-1}$) and $A_\text{ul}$ the Einstein coefficient for spontaneous emission,  which, in the case of a linear molecule such as HC$_3$,N can be obtained from Table~\ref{tab:hc3n_quants} using Eq.~25 from \citet{Goldsmith1999} and assuming an electric dipole moment of $\mu=3.73$\,D \citep{DeLeon1985}. Then, we relate upper level and total column densities by:
\begin{align}
    \label{eq:rdiag_0}
    \frac{N_u}{g_u}=\frac{N}{Z} e^{-E_u/kT}
\end{align}
where $Z$ is the HC$_3$N partition function obtained from the \texttt{CDMS}\footnote{\url{http://www.astro.uni-koeln.de/cgi-bin/cdmssearch}} catalogue \citep{CDMS1,CDMS2}. Rearranging and taking the logarithm on both sides of Eq.~\ref{eq:rdiag_0} we obtain:
\begin{align}
    \label{eq:rdiag}
    \log{\left(\frac{N_u}{g_u}\right)}=\log{\left(\frac{N}{Z}\right)}-\log{\left(e\right)} \cdot \frac{E_u}{kT}
\end{align}
from which we can derive the total column density $N$ and the excitation temperature $T$. by fitting a straight line using the least squares method. 

Figure~\ref{fig:rotdiag} upper and lower panels show the rotational diagram for the detected HC$_3$N* transitions toward the SHC and CND East condensation. 
The difference between the $T_\text{rot}$ and $T_\text{vib}$ is clearly illustrated by the dashed and color solid lines.
By fitting Eq.~\ref{eq:rdiag} to the transitions from the same rotational levels ($J=16-15$ or $J=24-23$) but different vibrational states, we estimate a vibrational temperature of $T_\text{vib,16-15}=239\pm30$\,K and $T_\text{vib,24-23}=390\pm122$\,K for the SHC.   These temperatures are clearly higher than the lower  rotational temperature $T_\text{rot,v=0}=103\pm14$\,K,  obtained by fitting Eq.~\ref{eq:rdiag} to all the transitions observed from the ground state (Fig.~\ref{fig:rotdiag}). For the CND East, since no HC$_3$N* was detected, we derived an upper limit $T_\text{vib,16-15} \leqslant 122$\,K from the $v_7=1$ $J=16-15$  and a $T_\text{rot,v=0}=42\pm10$\,K.  The errors have been obtained from bootstrapping using the uncertainties of the line fluxes. It is remarkable that the gas in the CND surrounding the SMBH seems to have lower excitation temperatures than the gas in the star-forming regions.

To fully account for the line profiles, opacity effects and line blending we have also carried out an LTE analysis of the HC$_3$N* emission using MADCUBA SLIM tool \citep{Madcuba2019}. Fig.~\ref{fig:SHC_spectra} and \ref{fig:CNDe_spectra} show the predicted SLIM line profiles superimposed  on the observed spectra for the SHC and CND East, respectively. The fitted parameters with their associated errors are given in Table~\ref{tab:table_hc3n_lums}. For the SLIM fitting we have used a source size of $0.42$\arcsec, the smallest beam with HC$_3$N data that reproduces the observed $v=0$ lines, although the emitting region of $v_7=1$ clearly must be smaller since it must arise from a region with a higher IR photon flux in order to be vibrationally excited.  We derive for the SHC $T_\text{vib}=236\pm16$\,K and $T_\text{rot}=98\pm7$,  similar to those derived from the rotational diagram. The derived line optical depths are $<0.01$, indicating that the optically thin assumption for the rotational diagram method is valid.
For the CND positions, where HC$_3$N* emission is not detected despite the large HC$_3$N $v=0$ column density, we have derived upper limits of $T_\text{vib}\lesssim 98$ and $\lesssim 117$\,K for the East and West positions, respectively. The derived $T_{\text{rot}}$  from the emission of the $v=0$ states in the CND is also much lower than in the SHC, reflecting lower excitation conditions (see below).

From the non-detection of HC$_3$N* in other star-forming regions of the SB pseudo-ring  we cannot completely rule out the presence of SHCs, since their upper limits to the emission from the $v_7=1$ lines are still consistent with  $T_\text{vib}\sim150-226$\,K as derived from $J=16-15$ HC$_3$N $v=0$ and $v_7=1$ lines upper limits. Furthermore, SSC 7 and 8 have  $T_{\text{rot},v=0}$ very similar to the SHC, indicating similar physical conditions that are fully consistent with the presence of SHCs, i.e. internally heated.  Conversely, SSC 6, 9 and 11  indicate  $T_\text{vib}\lesssim 180$\,K and  $T_{\text{rot}}$  $\sim 20-40$\,K, which are much lower than in the SHC, pointing to a lack of SHCs associated with these condensations. 

\begin{table*}
    \begin{center}
    \caption[]{Parameters derived from LTE and non-LTE modelling. LTE HC$_3$N column densities, rotational  temperatures from the ground state ($T_{\text{rot}}$) and vibrational temperatures  from  HC$_3$N* $J=16-15$ ($T_{\text{vib}}$) were obtained with SLIM assuming a source size of $0.42$\arcsec.  Positions marked with $^*$ have no HC$_3$N $v=0$ detected, their FWHM have been obtained from CS and a fiducial $T_{\text{rot}}=20\,K$ and $T_{\text{vib}}=100\,K$ were assumed in order to derive an upper limit for their HC$_3$N column densities. All non-LTE parameters are derived for a source size lower limit of $0.02$\arcsec.}
    \label{tab:table_hc3n_lums}
    \begin{threeparttable}
    \setlength{\tabcolsep}{3pt}
    
    \begin{tabular}{lcccccccccccc}
        \hline \noalign {\smallskip}
	            &	     \multicolumn{5}{c}{LTE}                   &	\, & \multicolumn{6}{c}{non-LTE}  \\ \cmidrule{2-6} \cmidrule{8-13}
	            & FWHM	& log N(HC$_3$N)$^\text{a}$& T$_{\text{rot}}$ 	&	T$_{\text{vib}}$  & L$_\text{LTE}^\text{b}$      &	& n$_{\text{H}_2}$&  log N(H$_2$)$^\text{c}$ &  log N(HC$_3$N)$^\text{c}$ & $X$(HC$_3$N) & T$_\text{d}$ & L$_\text{n-LTE}^\text{d}$  \\
Source	        &  (km\,s$^{-1}$)  &	(cm$^{-2}$) &	(K)   	&	(K)  &    ($10^8$\,L$_\odot$)   &\, &($10^5$\,cm$^{-3}$) &		(cm$^{-2}$) &		(cm$^{-2}$) & ($10^{-9}$) & (K) &    ($10^8$\,L$_\odot$)   	\\
        \hline \noalign {\smallskip}
SHC 	           & $33\pm2$ & $	14.7 (13.6)	        $ & $   98 \pm 7    $ & $	236 \pm 18	    $& $ 5.8\pm1.7$& & $5.9\pm0.2$ & $24.4(23.3)$ & $15.9(15.3)$ & $3.4\pm0.8$ & $248\pm28$ & $5.9\pm2.1$\\
SSC 2$^*$        & $32$ & $	\leqslant 13.2		    $ & $    20^* $ & $	100^*               $& $ \leqslant0.2$& &             &              &              &           &   \\
SSC 6            & $31\pm4$ & $	14.0 (12.9)		    $ & $   39 \pm 9    $ & $	\leqslant 150		$& $ \leqslant1.2$& & $3.0\pm0.1$ & $24.1(23.4)$ & $16.0(15.6)$ & $7.5\pm3.3$ & $\leqslant125$ & $\leqslant0.4$\\
SSC 7 	       & $27\pm4$ & $	14.1 (13.0)			$ & $   83 \pm 16   $ & $	\leqslant 213		$& $ \leqslant3.8$& & $3.1\pm0.6$ & $24.1(23.4)$ & $15.7(15.1)$ & $3.8\pm1.2$ & $\leqslant190$ & $\leqslant2.4$\\
SSC 8 	       & $26\pm7$ & $	14.0 (13.0)			$ & $   72 \pm 11   $ & $	\leqslant 226		$& $ \leqslant4.8$& & $5.9\pm0.1$ & $24.4(23.3)$ & $15.7(15.1)$ & $1.9\pm0.0$ & $\leqslant188$ & $\leqslant2.3$\\
SSC 9 	       & $31\pm2$ & $	13.7 (13.2)			$ & $   24 \pm 8    $ & $	\leqslant 178		$& $ \leqslant0.7$& & $1.5\pm1.0$ & $23.8(23.5)$ & $15.7(15.3)$ & $8.1\pm6.1$ & $\leqslant163$ & $\leqslant1.3$\\
SSC 11	       & $34\pm4$ & $	14.0 (13.1)			$ & $   36 \pm 7    $ & $	\leqslant 189		$& $ \leqslant2.4$& & $3.0\pm0.1$ & $24.1(23.7)$ & $16.0(15.6)$ & $7.1\pm3.2$ & $\leqslant181$ & $\leqslant2.1$\\
Pa\,$\alpha$ 1$^*$ & $31$ & $	\leqslant 13.3	        $ & $ 20 $ & $	100        $& $ \leqslant0.2$& &             &              &              &           &   \\
Pa\,$\alpha$ 2$^*$ & $30$ & $	\leqslant 13.4	        $ & $  20 $ & $	100		    $& $ \leqslant0.2$& &             &              &              &           &   \\
CND E.	           & $144\pm3$ & $ 15.0 (13.3)	        $ & $   41 \pm 1    $ & $	\leqslant 98	$& $ \leqslant0.2$& & $5.9\pm0.3$ & $24.4(23.3)$ & $16.0(15.5)$ & $3.7\pm1.1$ & $\leqslant75$ & $\leqslant0.1$\\
CND W.	           & $149\pm8$ & $ 14.4 (13.2)	        $ & $   29 \pm 3    $ & $	\leqslant 117	$& $ \leqslant0.3$& & $3.5\pm0.5$ & $24.2(23.6)$ & $15.7(14.8)$ & $3.0\pm0.6$ & $\leqslant114$ & $\leqslant0.1$\\
        \hline \noalign {\smallskip}
    \end{tabular}
        \begin{tablenotes}
      \small
      \item \textsuperscript{a} Column density derived from SLIM assuming a source size of  $0.42$\arcsec ($r=14.25$\,pc).
      \item \textsuperscript{b} Luminosities obtained by assuming a blackbody emitting at T$_{\text{vib}}$ and corrected from the back warming effect (see Sec.~\ref{subsec:proto_SSC_mass_and_lum}).
      \item \textsuperscript{c} Column density derived from the non-LTE models assuming a source size lower limit of  $0.02$\arcsec ($r=0.71$\,pc).
      \item \textsuperscript{d} Luminosities obtained from the non-LTE modelled SED between $10$ and $1200$\,$\mu$m  and corrected from the back warming effect.
    \end{tablenotes}
    \end{threeparttable}
    \end{center}
\end{table*}

\subsubsection{Non-LTE analysis}

To derive the physical properties of the SHC and the CND and to properly account for the different excitation mechanisms of the vibrational and rotational HC$_3$N transitions, we have carried out a non-LTE radiative transfer modelling using the same code used in \citet{Rico2020} \citep[described in detail in][]{GA97, GA99, GA19}, which includes the HC$_3$N rotational transitions up to $J=45$ in the  $v=0$, $v_7=1$ and $v_6=1$ vibrational states. As illustrated in Fig.\,4 by \citet{Rico2020}, the ratio between the $v_7=1$ and $v=0$ rotational lines from the same rotational level is extremely sensitive to the dust temperature, and the ratio between lines from different rotational levels but from the same vibrational state are dependent on the $n_{\text{H}_2}$ density. 

For NGC\,1068,  we  consider the ratios of the $v=0$ and $v_7=1$ $J=24-23$ and $J=16-15$ lines, since $J=39-38$ are undetected except for the SHC position. Using the ratio between the $J=16-15$ rotational transitions from the $v=0$ and $v_7=1$ states ($v_0/v_7$), we find for the SHC  a dust temperature of $T_{\text{d}}=248\pm28$\,K, very close to the previously derived $T_\text{vib}=236\pm18$,  and a density  $n_{\text{H}_2}=(5.9\pm0.2)\times10^{5}$\,cm$^{-3}$.

For the other positions, since no $v_7=1$ line is detected, we only have a lower limit on the ratio $v_0/v_7$ and we assume the $T_\text{vib}$ derived from LTE as their dust/kinetic temperature upper limit. To better constrain the dust temperatures and since the model also returns the SED, we also fit the observed continuum emission at $350$\,GHz. The obtained dust temperatures are similar to the $T_\text{vib}$  derived from the LTE modelling.

In particular, for the CND East position, the derived parameters from the models are
$T_{\text{d}}\leqslant 75$\,K;  a density  $n_{\text{H}_2}=(5.9\pm0.3)\times10^{5}$\,cm$^{-3}$,  and an HC$_3$N column density of $N_{\text{HC3N}}=(9.5\pm2.9)\times10^{15}$\,cm$^{-2}$ with a fractional abundance of $X_{\text{HC3N}}=(3.7\pm1.1)\times10^{-9}$ (Table~\ref{tab:table_hc3n_lums}). 
It is worth noting that the dust temperatures toward the CND East and West positions are much lower  than that derived for the SHC, despite the other parameters remaining similar to the SHC.  The parameters for the CND are in agreement with those estimated by \citet{Viti2014}, who derived densities for the CND of $(5-10) \times10^5$\,cm$^{-3}$ and a kinetic temperature of $60$\,K and $100-150$\,K for the CND East and West positions, respectively. \citet{GA14}, from modelling of the H$_2$O submillimeter emission,  found a $T_\text{d}\sim55$\,K and high densities $n(\text{H}_2)\sim10
^6$\,cm$^{-3}$, similar to the values we obtained.

\subsubsection{Proto-Super Star Cluster in the SB pseudo-ring:  mass and luminosity}
\label{subsec:proto_SSC_mass_and_lum}

The physical conditions of the SHC (and also SSC\,7 and 8) in NGC\,1068, high T$_\text{vib}$ and H$_2$ densities of few $10^5$\,cm$^{-3}$, are similar to those found in the Super Hot Cores (SHCs) of NGC\,253, where the IR emission from the dust heated by massive proto-stars vibrationally excites HC$_3$N. \citet{Rico2020} proposed that the SHCs trace the earliest phase of SSCs formation, the proto-SSC phase.  Following the analysis presented in  \citet{Rico2020}, we have estimated the LTE and non-LTE  luminosities of the condensations (clumps) studied in this work.  For the LTE luminosities we have used the emission of a black body with a (dust) temperature of  $T_{\text{vib}}$ and the lower limit to the size of the SHC of $0.021$\arcsec ($1.41$\,pc) derived by assuming that the emission from the $v_7=1$ HC$_3$N $J=16-15$ line is optically thick \citep[i.e. the source brightness temperature is equal to the derived vibrational/dust temperature; see][]{Rico2020}.  The same size is used to derive the upper limits for the other sources without detected HC$_3$N* emission. The non-LTE luminosity is derived from the spectral energy distribution between $10$ and $1200$\,$\mu$m predicted from the non-LTE models and assuming the same lower limit size as for the LTE estimate.

These estimates need to be corrected by the  back warming/greenhouse effect \citep{Donnison1976, GA19}, which appears in IR optically thick condensations  when a fraction of the IR radiation returns to the source (back warming), achieving the thermal equilibrium at a higher dust temperature than expected for the optically thin case. For the $N_{\text{H}_2}\sim 10^{24}$\,cm$^{-2}$ derived for the SHC, the apparent luminosities need to be corrected by a factor of $0.2$ \citep{Rico2020}, i.e. the actual luminosities will be $5$ times smaller than the directly estimated from the analysis, obtaining for the SHC $5.8\times10^8$\,L$_\odot$ and $5.9\times10^8$\,L$_\odot$ from the LTE and non-LTE models, respectively.  The  luminosities  corrected from back warming for all sources are listed in Table~\ref{tab:table_hc3n_lums}.

From these luminosities, we can make an estimate of the mass in proto-stars ($M_\text{p*}$) by assuming a luminosity-to-mass ratio of $10^3$\,L$_\odot$\,M$_\odot^{-1}$ \citep[similar to the luminosity-to-mass ratio typically assumed for ZAMS stars since the timescales for massive protostars to reach the ZAMS are short and  expected  to follow the ZAMS evolutionary track][]{Hosokawa2009, Rico2020}.  The proto-star masses of the SSCs in NGC\,1068 are given in Table~\ref{tab:summary_masses}.
The luminosity and mass of proto-stars in the SSCs of NGC\,1068 are similar to those found in NGC\,253. The non detection of HC$_3$N* emission in the remaining SSCs prevents us from firmly establishing to what extent they are still undergoing the proto-SSC phase (see below), apart from the SHC clump.

\section{Discussion}

\begin{table*}
    \begin{center}
    \caption[]{Summary table with all masses derived. The $M_{*,\text{Pa}\alpha}$ and $M_{*,147}$ are the ZAMS stellar masses derived from the Pa$\alpha$ and the $147$\,GHz continuum emission, respectively. $M_\text{gas}$ are the gas masses derived from the dust continuum emission at $350$\,GHz assuming the sizes obtained by fitting a two-dimensional Gaussian to each location, $\theta^2_{350}$ on Table~\ref{tab:continuum_fluxes}. $M_\text{p*,LTE}$ and $M_\text{p*,nLTE}$ are the proto stellar masses derived from the LTE and non-LTE modelling of HC$_3$N*. and $t_\text{age,p*}$ and $t_\text{age,Pa}$ are the SSC estimated ages following Eqs.~\ref{eq:tage} and \ref{eq:tage_pa}. SFE is the star formation efficiency.}
    \label{tab:summary_masses}
    \begin{tabular}{lrrrccrrrrc}
        \hline \noalign {\smallskip}
    Source	&	$M_{*,\text{Pa}\alpha}$	& 	$M_{*,147}$	&	$M_\text{gas}$	&	$M_\text{p*,LTE}$ & $M_\text{p*,nLTE}$	& $M_{*,147}/M_{*,\text{Pa}\alpha}$		&$M_\text{p*}/M_{*,147}$ & $t_\text{age,Pa}$ & $t_\text{age,p*}$ & SFE\\		
    	&	($10^5$\,M$_\odot$)	&	($10^5$\,M$_\odot$)	&	($10^5$\,M$_\odot$)	&($10^5$\,M$_\odot$)	    &	 ($10^5$\,M$_\odot$)&	&	& ($10^5$\,yr)& ($10^5$\,yr)	 & \\
    	\hline \noalign {\smallskip} 
SHC & $0.6$     & $7.9$ & $33.2$ & $5.8\pm1.7$      & $5.9\pm2.1$    & $13.0$   & $0.7$          &$	 0.6	$ & $	0.6	$ & $	0.2	$ \\
SSC 1  & $ 0.8 $ & $ 0.8 $ & $0.7$ &  &  & $ 1.0 $ &   &$	 \geqslant 100	$ &   & $	0.5	$ \\
SSC 2  & $1.4$   & $1.2$ & $3.8$  & $\leqslant0.2$   & $\leqslant0.2$  & $0.9$    & $\leqslant0.2$ & $	 \geqslant 100	$ & $\geqslant 	0.9	$ & $	0.2	$\\
SSC 3  & $ 0.4 $ & $ 0.9 $ & $1.0$ &  &  & $ 2.6 $ &  &$	 4.5	$ & & $	0.5	$ \\
SSC 4  & $ 0.8 $ & $ 1.4 $ & $2.8$ &  &  & $ 1.7 $ &  &$	 9.7	$ & & $	0.3	$ \\
SSC 5  & $ 0.3 $ & $ 2.0 $ & $8.2$ &  &  & $ 6.6 $ &  &$	 1.3	$ & & $	0.2	$ \\
SSC 6  & $0.3$   & $4.1$ & $9.3$  & $\leqslant1.2$   & $\leqslant0.4$ & $13.5$   & $\leqslant0.3$ &$	 0.6	$ & $\geqslant 	0.8	$ & $	0.3	$\\
SSC 7  & $1.5$   & $4.4$ & $10.5$ & $\leqslant3.8$   & $\leqslant2.4$ & $2.9$    & $\leqslant0.7$ &$	 3.8	$ & $\geqslant 	0.5	$ & $	0.3	$\\
SSC 8  & $1.4$   & $4.2$ & $18.0$ & $\leqslant4.8$   & $\leqslant2.3$ & $2.9$    & $\leqslant1.2$ &$	 3.8	$ & $\geqslant 	0.5	$ & $	0.2	$\\
SSC 9  & $0.4$   & $2.3$ & $15.2$ & $\leqslant0.7$   & $\leqslant1.3$ & $5.1$    & $\leqslant0.3$ &$	 1.8	$ & $\geqslant 	0.8	$ & $	0.1	$\\
SSC 10 & $ 0.7 $ & $ 1.4 $ & $3.0$ &  &  & $ 2.1 $ &  &$	 6.8	$ & & $	0.3	$ \\
SSC 11 & $0.5$  & $2.4$ & $4.0$  & $\leqslant2.4$   & $\leqslant2.1$ & $4.7$    & $\leqslant0.2$  &$	 2.0	$ & $\geqslant 	0.5	$ & $	0.4	$\\
SSC 12 & $ 0.5 $ & $ 1.2 $ & $1.9$ &  &  & $ 2.3 $ &  &$	 5.8	$ & & $	0.4	$ \\
SSC 13 & $ 0.7 $ & $ 1.1 $ & $3.1$ &  &  & $ 1.5 $ &  &$	 12.9	$ & & $	0.3	$ \\
SSC 14 & $ 0.3 $ & $ 1.0 $ & $5.0$ &  &  & $ 3.7 $ &  &$	 2.7	$ & & $	0.2	$\\
Pa\,$\alpha$ 1 & $0.9$     & $1.0$ & $\leqslant0.5$ & $\leqslant0.2$ &  & $1.1$ & $\leqslant0.2$ &$	 40.6	$ & $\geqslant 	0.8	$ & $	0.7	$\\
Pa\,$\alpha$ 2 & $0.6$     & $\leqslant0.8$ & $\leqslant0.2$ & $\leqslant0.2$ &  & $\leqslant1.3$& $\leqslant0.2$ &$\geqslant	 21.2	$ & $	\geqslant 0.8	$ & $	0.8	$\\
            \hline \noalign {\smallskip}
    \end{tabular}
    \end{center}
\end{table*}

\subsection{On the heating of the SHC and the CND}
\label{sec:heating_CND}

As already mentioned, the excitation of HC$_3$N  can be dominated by two different mechanisms: IR radiation pumping by hot dust and/or  collisions with H$_2$, which can be used to discriminate between different heating mechanisms. We have found a significant difference between the derived $T_\text{vib}$ for the CND East and West positions ($\lesssim 100$\,K) and for the SHC ($\simeq 240$\,K). 
The same trend is also found for the rotational temperature: T$_{\text{rot}}= 41$\,K for the CND East, $29$\,K for CND West and $98$\,K for the SHC, with SSCs 7 and 8 having also high T$_{\text{rot}}\sim80$\,K.  The derived H$_2$ densities for both the SHC and the CND positions are similar, but much lower than the critical density for collisional excitation of the vibrational levels \citep[$>10^8$\,cm$^{-3}$,][]{Wyrowski1999}. Therefore, the excitation of the vibrational states must be through IR pumping by the hot dust and the excitation of the rotational levels by collisions with H$_2$.

 The CND, located at $\sim60-80$\,pc from the SMBH in NGC\,1068, seems to be strongly affected by the interaction with the AGN. Both radiative and mechanical effects have substantially changed its  kinematics and physical and chemical properties \citep{Tacconi1994, Usero2004, SGB2010}.  X-rays from the AGN have been proposed to explain the specific chemistry found in the CND \citep[e.g.][]{Sternberg1994, Usero2004, Aladro2012, Aladro2013}.
Several studies have analyzed the influence of the AGN on the physical properties and chemical composition of the CND.  Most of them disregard the effects of the UV radiation and consider shocks and/or X-rays irradiation to be the most plausible  mechanisms  heating the CND \citep[e.g.][]{Galliano2002, Usero2004, Krips2011, Hailey2012, Spinoglio2012, Aladro2013, SGB2014, Viti2014}.
Based on  chemical modelling, \citet{Viti2014}, have proposed that the CND can be characterized by a three-phase component ISM: two components with enhanced cosmic-ray ionization rates by a factor of $10$ compared to the Milky Way, and/or X-rays, but with different densities ($10^5$ and $\geqslant 10^6$\,cm$^{-3}$);  and a third component dominated by shocks from the outflow driven by the AGN. Our derived  H$_2$ density of $6\times10^5$\,cm$^{-3}$ for the CND East and $4\times10^5$\,cm$^{-3}$ for the CND West, are in agreement with the densities derived by \citet{Viti2014}. 

We have found significant lower vibrational temperatures for the CND East and West positions ($T_\text{vib}\lesssim 75$\,K and $\lesssim 114$\,K)  than the gas kinetic temperatures derived by  \citet{Viti2014}  ($80-160$\,K  and $>100$\,K)  and than the vibrational temperature obtained for the SHC ($T_\text{vib}\sim 240$\,K).
Furthermore, the rotational temperatures of the CND East and West positions derived from the ground state ($49$\,K and $21$\,K) are also lower than the kinetic temperatures ($T_\text{rot}$) of the SHC , which is  an indication of sub-thermal collisional excitation.

 As discussed in Sec.~\ref{subsec:derivedHC3Nprops}, this clearly reflects that the IR pumping is not efficiently exciting HC$_3$N* on the CND. This could be due to a low dust optical depth at the wavelength of the vibrationally transitions ($45$\,$\mu$m for the $v_7=1$) or due to a low dust temperature, both effects indicate that the flux of IR photons being re-emitted (i.e. trapping) by dust at $45$\,$\mu$m is small and thus not enough to excite HC$_3$N*. The first possibility can be ruled out since the H$_2$ column densities in the CND are $\sim10^{24}$\,cm$^{-2}$ (see Table
~\ref{tab:table_hc3n_lums}), which translates to dust opacities at $45$\,$\mu$m of $\tau \sim16$. Then, the other option is a low dust temperature in the CND in spite of the large  luminosity  estimated from mid-IR   $L_\text{AGN}\simeq1.1\times10^{11}$\,L$_\odot$ \citep{Alonso-Herrero2011, SGB2014}. This indicates that  the dust in the CND remains at a low temperature $<100$\,K \citep{SGB2014}, evidencing that the AGN is not effective at heating the dust in its surroundings. Indeed, as shown below, the upper limit to the dust temperature is consistent with the  expected temperature for the heating from the AGN when considering that the dust in the CND is being heated externally. Considering that the CND is a condensation detached from the AGN located at  a projected distance from the AGN of $\sim75$\,pc, the expected dust temperature for external heating will be $57$\,K taking into account the heating for the IR optically thin case using the Stefan-Boltzmann law \citep{deVicente2000}:
 \begin{align}
 \label{eq:tdust}
 T_\text{d}(\text{K})=\left(\frac{L}{4\pi\sigma r^2}\right)^{1/4}=15.222\left(L(\text{L}_\odot)\left[\frac{10^{16}}{r(\text{cm})}\right]^2\right)^{1/4}
 \end{align}
 
This indicates that the dust in the CND not being efficiently heated by the AGN is due to being externally irradiated by the AGN (i.e. geometrical open system), where photon trapping in the IR is negligible and the temperature profile follows effectively the optically thin case. The opposite happens in geometrical closed systems, where the source is being internally irradiated and is optically thick in the IR. In this case, the greenhouse effect dramatically changes the dust temperature profile in the surrounding material, as observed in the SHC.

The amount of  HC$_3$N* found in NGC\,253 \citep{Rico2020} and that found in the NGC\,1068 SHC location, contrasts with its absence in the CND of NGC\,1068, where HC$_3$N column densities are similar to those found in the SHC. 
While in NGC\,253 UV radiation has been found to be the dominant excitation mechanism of H$_2$ emission \citep{Rosenberg2013}, in the CND of  NGC\,1068 it is most likely to be heated by X-ray irradiation \citep{Galliano2002}, where gas is also being heated by shocks \citep{Kelly2017, SGB2014, Viti2014}.

\subsection{History of star formation in the SB pseudo-ring of NGC\,1068}

\subsubsection{Ages of SSCs}

We can study the recent history of massive star formation in the SB pseudo-ring of NGC\,1068 by comparing the different tracers presented in this work. As discussed in \citet{Rico2020}, the  HC$_3$N* emission is thought to be tracing the proto-SSCs phase. 
Then the free-free radio continuum emission at $147$\,GHz, arising from Ultra Compact \ion{H}{II} regions, would be tracing an early phase in the evolution of massive ZAMS stars just after the proto-SSC stage, as seen by their $\alpha_{350-150}>1$ indexes.
And finally, the  Pa$\alpha$ emission will be tracing a more evolved phase of  massive young stars in the SSCs, as indicated by the lower level of extinction associated to its surrounding material, which is being removed by stellar feedback. In fact, the ratio between the SSCs star masses estimated from $147$\,GHz  and from the Pa$\alpha$ emission  ($M_{*,147}/M_{*,\text{Pa}\alpha}$)  can be used to make a rough estimate of the evolutionary stage of the SSCs. The $M_{*,147}/M_{*,\text{Pa}\alpha}$ mass ratio is expected to be close to $1$ for no extinction, and it will increase as the extinction increases, since with the continuum emission at $147$\,GHz we are seeing a more embedded phase of star formation.
This is indeed the trend  found in the $M_{*,147}/M_{*,\text{Pa}\alpha}$ ratio  in  Table~\ref{tab:summary_masses}.  We can see that the SSCs with the lowest values of about $1$, SSC\,1, SSC\,2, Pa$\alpha$ 1 and Pa$\alpha$ 2, almost do not show dust continuum emission at $350$\,GHz, as expected for low extinction. 
It is worth to remark that we are focusing on specific clumps in the SB pseudo-ring where the dust continuum emission at $350$\,GHz peaks (except for the Pa$\alpha$\,1 and 2 sources) and hence the Pa$\alpha$ emission is expected to suffer some extinction, although it seems to have no extinction outside these clumps (for a more detailed analysis on the Pa$\alpha$ emission in the SB pseudo-ring see S\'anchez-Garc\'ia et al. in prep.).
The highest values of the $M_{*,147}/M_{*,\text{Pa}\alpha}$ ratios correspond to the proto-SSC (i.e. the SHC) and the SSC\,6 located in the northern region of the SB pseudo-ring. The high value found in the proto-SSC is consistent with the very early stage of the SSC formation.     Following the procedure described by  \citet{Rico2020}, we can use the ratio of the mass in proto-stars ($M_\text{p*}$) to the the mass in ZAMS stars ($M_{*,147}$), to make an estimation of their age assuming that proto-stars have a timescale of $\lesssim 10^5$\,yr with:
\begin{align}
\label{eq:tage}
t_\text{age}(\text{yr}) \sim
\frac{1}{1+M_{\text{p*}}/M_\text{*,147}} \times 10^5 
\end{align}

The estimated age of the proto-SSC is $5.8\times10^4$\,yr. For the more evolved SSCs without detected HC$_3$N* emission, we can estimate their ages  using the ratio  $M_{*,147}/M_{*,\text{Pa}\alpha}$.
Assuming that the feedback from the SSCs removes the leftover material from cluster formation in $\sim10^7$\,yr \citep{Dowell2008} and assuming a linear dependence of the gas removal with age, we can estimate the age as:
\begin{align}
\label{eq:tage_pa}
t_\text{age,Pa} \text{(yr)}= \frac{1}{1+13\cdot(M_{*,147}/M_{*,\text{Pa}\alpha}-1)}\times10^7
\end{align}
where $13$ is a scale factor obtained from the SHC position so its $t_\text{age,p*}=t_\text{age,Pa}$.
The ages obtained are listed on Table~\ref{tab:summary_masses}. We find that the youngest SSCs are the SHC and SSC\,6 (ages of $6\times10^4$ years) located in the northern part of the SB pseudo-ring, and SSC\,7, 8, 9 and 11 (ages $2-4\times10^5$ years) in the southern part of the SB pseudo-ring. It is remarkable that the SSCs seems to be  closely associated with the nuclear stellar bar and the beginning of the molecular spiral arms.

We can use the gas mass obtained from the $350$\,GHz continuum emission to make an estimation of the Star Formation Efficiency (SFE, Table~\ref{tab:summary_masses}) by assuming that the initial mass of the star-forming molecular clouds has not suffered significant mass losses. Since we only have  upper limits for $M_\text{p*}$ for most of the SSCs, we will use only $M_{*,147}$ and the expression:
\begin{align}
    SFE = \frac{1}{1+M_\text{gas}/M_{*,147}}
\end{align}
In case that mechanical feedback has played a significant role, like in the case of Pa$\alpha$ 1 and 2, the SFE must be considered as an upper limit. 

The SFE of the proto-SSC is only $0.2$, suggesting that the star formation is very recent and still has a significant amount of gas available to convert into stars. SSCs\,6, 7, 8, 9 and 11, with ages of few $10^5$ yr, seem to be in a similar state to the SHC, where the latest star formation episode is just starting (or has just concluded). Unfortunately the lack of sensitivity to detect HC$_3$N* does not allow to discriminate between the two possibilities. SSC\,2 shows a relatively high gas mass and a relatively low SFE, but seems to be rather evolved as its $M_{*,147}/M_{*,\text{Pa}\alpha}$ ratio suggests.  This could indicate that its star formation has been halted and a next generation of stars could be forming in an unrelated giant molecular cloud unresolved by our observations. On the other hand, SSC \,1 has a SFE$\lesssim0.5$ with low M$_\text{gas}$, indicating that it has already formed most of its stars.

\begin{figure*}
\centering
    \includegraphics[width=\linewidth]{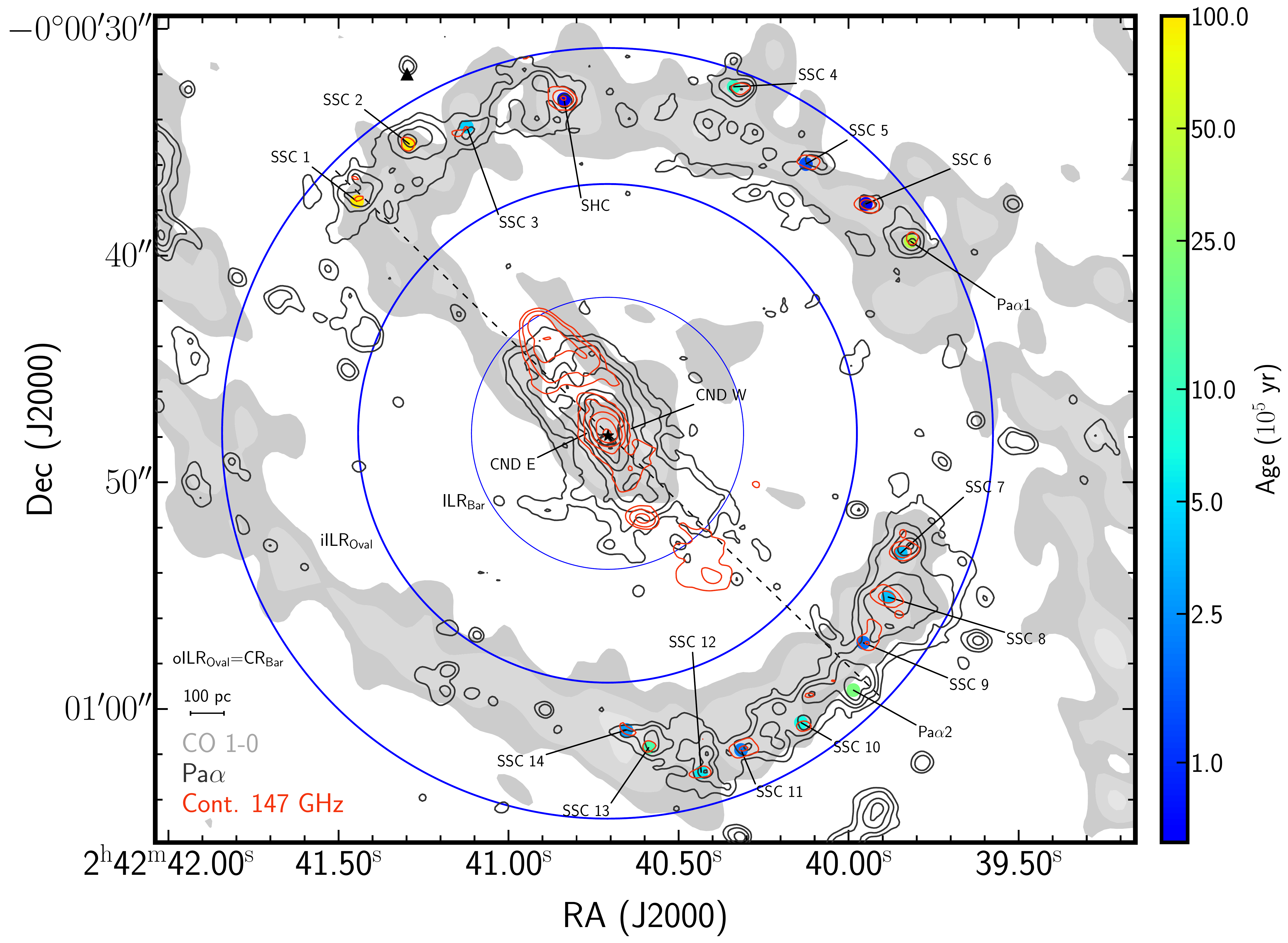}
  \caption{NGC\,1068 map as seen by CO $1-0$ (grey scale), Pa\,$\alpha$ (black contours) and continuum emission at 147\,GHz (red contours). The SB pseudo-ring is clearly seen in CO and Pa$\alpha$ emission. The AGN position is marked with a star. Type II SN 2018ivc position is marked with a triangle \citep{Bostroem2020}. The stellar bar  is indicated with a straight dashed black line. The ILRs from the outer oval and the stellar bar \citep{Schinnerer2000, SGB2010} are indicated by solid blue lines, respectively. The estimated ages ($t_\text{Age,Pa}$) for the different positions are color coded following the color wedge on the right.
  }
  \label{fig:CO}
\end{figure*}

\subsubsection{Propagation of SSC formation in the ring}

NGC\,1068 is a classical case of two embedded bars that have resonances in common. The large-scale stellar oval of length $\sim17$\,kpc \citep{Telesco1988, Schinnerer2000}  and the nuclear stellar bar of $r\sim1.3$\,kpc and a position angle $46^\circ$;]\citep[][]{Scoville1988, Schinnerer2000}. The large-scale oval has its corotation at $R_\text{cor}=120\arcsec-140\arcsec \sim 8.5-10$\,kpc \citep[see][]{Schinnerer2000}, while the nuclear stellar bar corotation is at $R_\text{cor}=15\arcsec-2\arcsec \sim 1-1.5$\,kpc. This corotation of the nuclear bar coincides with the outer Inner Lindblad Resonance (oILR) of the large-scale oval. These overlapping resonances favour the decoupling of the nuclear bar from the large-scale bar. There is a tightly-wound nuclear spiral in the gas (\quotes{the SB pseudo-ring}) that is identified by the accumulation of molecular gas (seen in CO in Fig.~\ref{fig:CO}) in a pseudo-ring at radii $r=15-20\arcsec$, i.e. at the corotation of the nuclear bar. The formation of a pseudo-ring is a natural consequence of the gas responding to the dynamical resonances induced by a bar-shaped potential.The CO arms are at the nuclear bar corotation radius indicating that the gas is responding to the larger-scale oval in NGC\,1068, which would favour inflows to its ILR. \citet{SGB2014}.

Assuming that gas rotation in NGC\,1068  is counter-clockwise if the spiral arms are trailing, the CO ridge appears \quotes{upstream} (i.e. at shorter radii, on the inner face of the spiral arm) along the expected gas circulation lines while the star formation tracers (both Pa\,$\alpha$ and $147$\,GHz Continuum) tend to appear \quotes{downstream} (i.e. at longer radii, on the outer face of the spiral arm). This is the expected \quotes{sequence} if the gas is responding to the spiral inside corotation: it enters the spiral arm, where is being accumulated and/or compressed and then star formation is triggered. An example of this sequence is seen in M\,51 spiral arms \citep[][]{Louie2013}.
If we zoom in the SB pseudo-ring, one would also expect to see the older star formation (i.e. the Pa\,$\alpha$ emission) downstream relative to the younger and still embedded star formation (i.e. the $147$\,GHz Continuum and/or the HC$_3$N*). Hints of this intricate \quotes{sub-sequence} can be found on the SW region but is less clear on the NE region.

A sizeable fraction of the massive recent star formation is concentrated in two regions located NE and SW along $\text{PA}=46^\circ$, where the CO spiral arms join with the end of the nuclear bar (see Fig.~\ref{fig:CO}). These are the regions where \citep{SGB2014} identified strong non-circular motions revealing an inflow and where the orbit families of the spiral and the bar are expected to cross, which would favour cloud-cloud collisions that enhance the star formation.
\citet{Beuther2017} observed a similar built of resonances in NGC\,3627 that pile up the gas where the different orbit families cross, favouring strong star formation events. This is also occurring in the Milky Way, where the overlap between the end of the Galactic bar with the spiral arm hosting the W43 starburst region \citep[see][and references therein]{Beuther2017}.

\begin{figure}
	\includegraphics[width=0.9\linewidth]{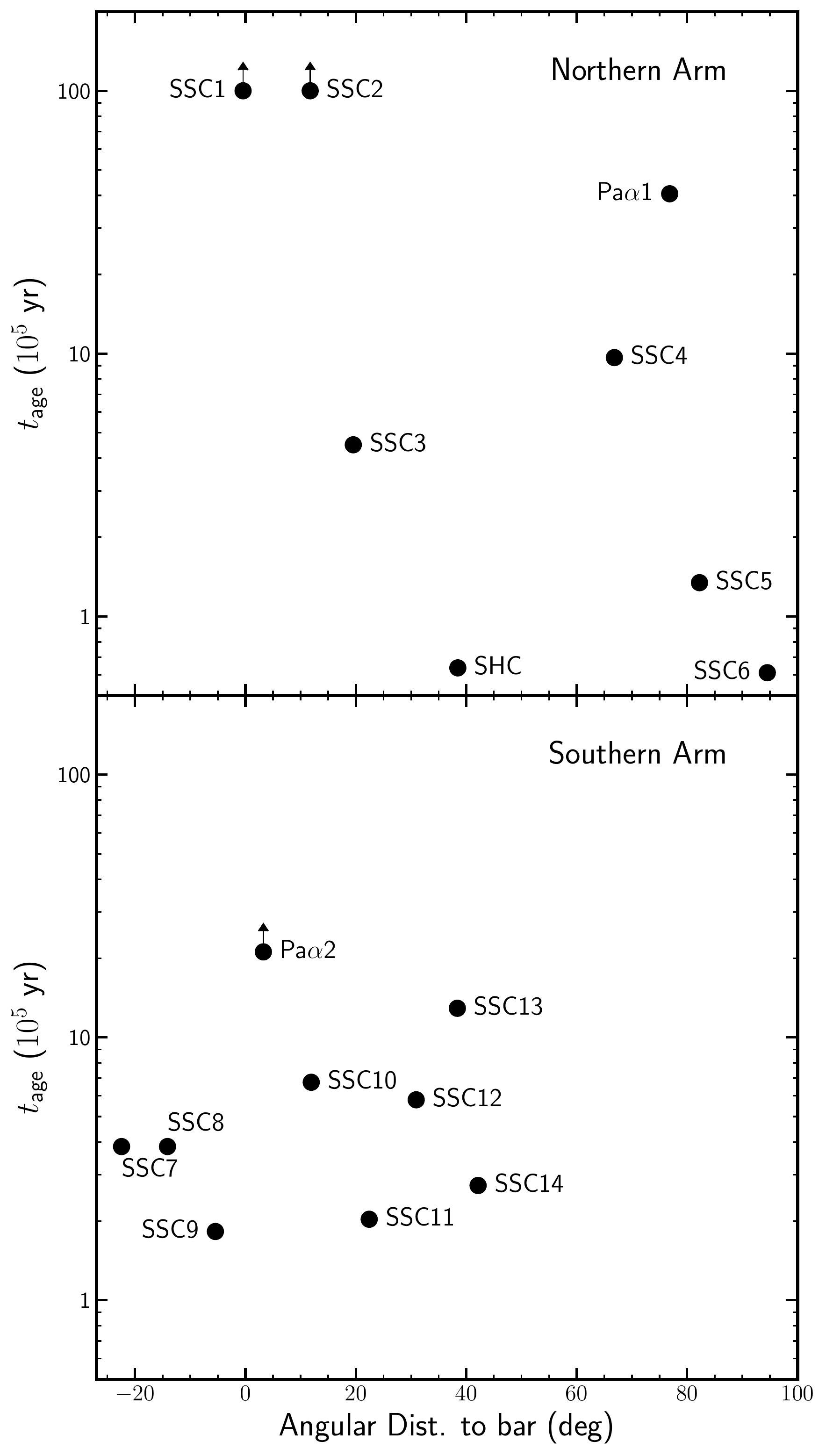}
    \caption{SSCs ages as a function of their relative distance to the nuclear bar. Top panel shows the SSCs on the northern spiral arm. Bottom panel shows the SSCs on the southern spiral arm.}
    \label{fig:ang_pos}
\end{figure}

Despite the inflow gas motions that lead to the molecular gas pile-up in the spiral arms, star formation is not uniformly distributed throughout the entire SB pseudo-ring: the two contact-point regions located SW and NE stand out as the sites where the bulk of the recent and massive SF is taking place. However, if we consider the different SSCs ages  (from Table~\ref{tab:summary_masses}) ant their projected location relative to the nuclear bar (Figure~\ref{fig:ang_pos}), it seems that there is no ordered sequence of the star forming regions inside these two regions. 
This would be in accordance with the \quotes{popcorn} scenario instead of the \quotes{pearls on a string} scenario described by \citet{Boker2008}. In the \quotes{popcorn} scenario, the star forming regions would appear at random times and locations within the pseudo-ring and there would be no systematic sequence of star formation along the pseudo-ring. The collection of different initial conditions of each cloud when they enter the spiral arm would make the triggering of the star formation appear at random times.

\subsection{HC$_3$N* as a discriminator between AGN and star formation activity in galaxies}

The different excitation mechanism present in the CND, where no recent star formation is taking place \citep[last starburst was $200-300$\,Myr ago][]{Davies2007}, compared to those observed in the young star-forming regions where HC$_3$N* has been observed, suggests that HC$_3$N* could be used as a discriminator between AGN and early obscured star formation activity in galaxies.

So far, HC$_3$N* has been detected in two active galaxies, the LIRG NGC\,4418, located at $35.8$\,Mpc with $L_\text{IR}=1.5\times10^{11}$\,L$_\odot$ \citep{Costagliola2010}, and the ULIRG Arp\,220 \citep{Martin2011}, located at $79.4$\,Mpc and with $L_\text{IR}=1.5\times10^{12}$\,L$_\odot$. For both galaxies the origin of the high IR luminosity (AGN versus SB) is highly debated due to the extremely large extinctions. For NGC\,4418, \citet{Sakamoto2013b} favored  the heating  by a Compton-thick AGN, but \citet{Varenius2014} claimed to be SB dominated. A similar controversy also exists for Arp\,220, \citep[e.g.][]{Sakamoto1999, Wilson2014, Barcos2015, Sakamoto2017}.

Following the results found for the CND we can compare the estimated $T_\text{vib}$, HC$_3$N* brightness temperature and source sizes for NGC\,4418 and Arp\,220. Using results from previous works, we can compare between the sizes derived from HC$_3$N* emission and those predicted from Eq.~\ref{eq:tdust} to discriminate between obscured AGN and SF scenarios.
For NGC\,4418,  \citet{Costagliola2015} estimated a $T_\text{vib}=340$\,K, with a brightness temperature in the $J=24-23$, $v_7=1f$ line of $9$\,K for a source size of $0.4\arcsec$ ($69$\,pc). Assuming optically thick emission for that line, we derive a lower limit to the size of $\sim0.07\arcsec$ ($\sim12$\,pc). For Arp\,220, \citet{Martin2011} measured for the $J=25-24$ $v_7=1f$ line a  mean beam brightness temperature of $263.9$\,mK in a $8.4\arcsec \times 6.1\arcsec$ beam and estimated a $T_\text{vib}\sim355$\,K. Applying the same procedure as for NGC\,4418, we derive a lower limit to the HC$_3$N* emission of $\sim0.2\arcsec$ ($\sim80$\,pc) for Arp\,220.

Taking into account the results obtained for the heating of the CND by the AGN in NGC\,1068 (Eq.~\ref{eq:tdust}), one would expect the size of the region emitting the bulk of the hot gas ($\sim 340$\,K) observed in  HC$_3$N* for NGC\,4418 and Arp\,220, to be of about 
$2.5$ and $8.0$\,pc respectively (i.e. $0.015\arcsec$ and $0.021\arcsec$), if the heating is dominated by a central source as expected for an AGN.
The predicted sizes by Eq.~\ref{eq:tdust} are clearly much smaller than the measured lower limits to the sizes of the HC$_3$N* emission in both sources.
The rather small  predicted sizes for the dust region heated by the AGN would indicate that the heating is instead distributed over  the large region observed in HC$_3$N* due to star formation as observed in the NGC\,1068 SB pseudo-ring.

However, as discussed in Sec.~\ref{sec:heating_CND}, Eq.~\ref{eq:tdust} can only be applied to predict the dust temperature profile for the case of an optically thin cloud (open systems) in the IR. 
 \citet{GA19} have studied in detail the dust temperature profile due to the heating by an AGN and a SB in a spherical cloud considering the back warming effect due to extremely high extinction (i.e. closed systems) as observed in NGC\,4418 and Arp\,220. They found that for a luminosity surface brightness  of $1-2\times10^8$\,L$_\odot$\,pc$^{-2}$ for NGC\,4418 and Arp\,220 and an  H$_2$ column density of $10^{25}$\,cm$^{-2}$, the size of the region with $T_\text{d} > 300$\,K is of about $20$\,pc and $100$\,pc for NGC\,4418 and Arp\,220, basically independent from the nature of the heating source. Therefore, discriminating between AGN and SB heating using HC$_3$N* emission in IR optically thick galaxy nuclei can only be done when the spatial resolution is high enough to measure the $T_\text{vib}$ profile close to the central heating source, where the temperature gradient is expected to show the largest difference between the AGN and SB heating.

\section{Summary and conclusions}

We have used archival ALMA data to study the HC$_3$N emission from its ground and vibrationally excited states, along with the $147$\,GHz continuum emission, in the SB pseudo-ring and in the CND of NGC\,1068. The main results can be summarised as follows:

\begin{itemize}

\item[--]We have detected emission from HC$_3$N (lines $J=11-10$, $12-11$, $16-15$, $24-23$)  in the ground state towards the SB pseudo-ring and the CND. In spite of the bright HC$_3$N emission observed towards the CND East, we did not detect any vibrationally excited emission. In contrast, vibrationally excited emission from the $v7=1e$ and $1f$ lines ($J=16-15$ and $24-23$) of HC$_3$N was detected towards one star-forming region on the northern part of the SB pseudo-ring.

\item[--] For the star-forming region in the SB pseudo-ring with HC$_3$N* emission, the LTE analysis  yields a vibrational temperature ($T_\text{vib}$) between the $v=0$ and $v_7=1$ levels of $236\pm18$\,K and a rotational temperature, $T_\text{rot}$,  between the rotational levels in the ground state of $98\pm7$\,K. The difference in excitation temperatures suggests that the vibrational levels are excited by IR pumping at the dust temperature  while the rotational levels are  collisionally excited.  This is consistent with the derived $T_\text{dust}$ and H$_2$ densities from our non-LTE analysis of $248\pm28$\,K and $(5.9\pm0.2)\times10
^5$\,cm$^{-3}$, respectively. The latter indicates that we are observing a Super Hot Core (SHC) similar to those observed in NGC\,253 \citep{Rico2020}. 

\item[--] From the dust temperature and the lower limit to the size of the SHC in the SB pseudo-ring we estimated an IR luminosity of $\sim5.8\times10^8$\,L$_\odot$, typical of the proto-Stellar Star Clusters (proto-SSC) observed in NGC\,253, which are believed to be tracing the earliest phase of the SSC formation.

\item[--]  In addition to the SHC, we have also identified from our continuum map at $147$\,GHz another $14$ young star-forming regions undergoing the \ion{H}{II} region phase in the SB pseudo-ring. Assuming that the continuum is dominated by free-free emission, we have obtained the stellar mass of ZAMS massive stars for the regions in the SB pseudo-ring. These embedded star-forming regions contain stellar masses $\sim10^5$\,M$_\odot$, typical of SSCs and proto-SSC. Unfortunately, our sensitivity on the HC$_3$N* emission is not high enough to discard the presence of the proto-SSC phase.

 \item[--] We have also used the Pa$\alpha$ emission to trace a more evolved star formation phase in the SB pseudo-ring. We  combined the tracers of the different evolutionary phases to estimate the ages of the SSCs, which range from few 10$^4$\,yr for the proto-SSCs  to 10$^7$\,yr for the most evolved SSCs with little extinction as derived from the ratio of the Pa$\alpha$ to $147$\,GHz derived stellar masses and the lack of dust/CO emission. 
 
 \item[--] The SSCs  location and derived young ages are closely associated to the region connecting the nuclear bar with the SB pseudo-ring, where most of the gas is being accumulated and an increase of cloud-cloud collisions is expected. The most recent and still embedded star formation episodes would be taking place in this region, as expected in the scenario of the inflow scenario. We do not find any systematic trend of the ages of the SSCs within the ring, supporting the \quotes{popcorn} mode of star formation.

\item[--] For the CND East and West positions, our LTE and non-LTE analysis of  HC$_3$N emission yield an upper limit to $T_\text{vib}$ of $\leqslant98$\,K and $\leqslant117$\,K and a $T_\text{rot}$ of $41\pm1$\,K and $29\pm3$\,K, respectively. 
The derived H$_2$ densities are of  $(5.9\pm0.3)\times10
^5$ and $(3.5\pm0.5)\times10
^5$\,cm$^{-3}$.

\item[--] The low dust temperature $<100$\,K derived towards the CND, is consistent with the expected heating by the AGN in NGC\,1068 for a luminosity of $1.1\times10^{11}$\,L$_\odot$ considering the IR optically thin case and that the dust in the CND is being externally heated (open system). The opposite happens in the SHC, which is being internally irradiated by the forming proto-stars and is optically thick in the IR (closed system).

\item[--] We discussed if, as observed in NGC\,1068, the HC$_3$N* emission observed in NGC\,4418 and Arp\,220 can be used to discriminate between AGN and SB activity. We concluded that just the detection of HC$_3$N* emission cannot be used as a discriminator because of the greenhouse effect in  heavily obscured galactic nuclei (closed systems) makes the AGN and SB dust profiles to be similar at large distances. Only the combination of spatially resolved images of several HC$_3$N* lines might provide the insight for the discrimination.
    
\end{itemize}

\section*{Acknowledgements}

This paper makes use of the ALMA data listed in Table~\ref{tab:observations}. ALMA is a partnership of ESO (representing its member states), NSF (USA) and NINS (Japan), together with NRC (Canada) and NSC and ASIAA (Taiwan) and KASI (Republic of Korea), in cooperation with the Republic of Chile. The Joint ALMA Observatory is operated by ESO, AUI/NRAO and NAOJ.  

The Spanish Ministry of Science and Innovation has supported this research under grant number ESP2017-86582-C4-1-R, PhD fellowship BES-2016-078808 and MDM-2017-0737 Unidad de Excelenc\'ia Mar\'ia de Maeztu.

V.M.R. acknowledges support from the Comunidad de Madrid through the Atracci\'on de Talento Investigador Senior Grant (COOL: Cosmic Origins Of Life; 2019-T1/TIC-15379), and from the  European Union's Horizon 2020 research and innovation programme under the Marie Sk\l{}odowska-Curie grant agreement No 664931.

I.J.-S. has received partial support from the Spanish FEDER (project number ESP2017-86582-C4-1-R) and the State Research Agency (AEI; project number PID2019-105552RB-C41).

\section*{Data Availability}

The data underlying this article were accessed from the ALMA Science Archive (\url{http://almascience.eso.org/asax/}), with the corresponding project identifiers listed on Table~\ref{tab:observations}. The derived data generated in this research will be shared on reasonable request to the corresponding author.



\bibliographystyle{mnras}
\bibliography{NGC1068.bib}



\bsp	
\label{lastpage}
\end{document}